\documentclass[journal,twoside]{IEEEtran}
\usepackage{bm}

\usepackage{tabularx}
\usepackage{booktabs}
\usepackage{multirow}

\usepackage{cite}

\ifCLASSINFOpdf
  \usepackage[]{graphicx}
  \graphicspath{{./Fig/}{.}}
  \DeclareGraphicsExtensions{.pdf,.jpeg,.png}
\else
  \usepackage[]{graphicx}
  \graphicspath{{./Fig/}}
  \DeclareGraphicsExtensions{.eps}
\fi
\usepackage[cmex10]{amsmath}
\usepackage{algorithmic}
\usepackage{algorithm}

\usepackage{array}

\ifCLASSOPTIONcompsoc
 \usepackage[caption=false,font=normalsize,labelfont=sf,textfont=sf]{subfig}
\else
 \usepackage[caption=false,font=footnotesize]{subfig}
\fi
\begin{document}
%
\title{Bayesian full waveform inversion with learned prior using deep convolutional autoencoder}
%
%
%

\ifCLASSOPTIONpeerreview


\else

\author{Shuhua~Hu,~\IEEEmembership{}
        Mrinal~K Sen,~\IEEEmembership{}
        Zeyu~Zhao,~\IEEEmembership{Member,~IEEE,}
        Abdelrahman~Elmeliegy,~\IEEEmembership{Member,~IEEE,}
        and~Shuo~Zhang~\IEEEmembership{}
\thanks{S. Hu, M. Sen, A. Elmeliegy, and S. Zhang are with the University of Texas at Austin, Jackson School of Geosciences, Institute for Geophysics, Department of Earth
and Planetary Science, Austin, TX 78758 e-mails: sh.hu@utexas.edu, mrinal@utexas.edu, abdelrahman.elmeliegy@inl.gov, shuo.zhang@jsg.utexas.edu; S. Hu is currently with Halliburton Landmark, Houston, TX 77032; A. Elmeliegy is currently with Idaho National Laboratory, Idaho Falls, ID 83415; S. Zhang is currently with School of Earth Sciences, Yunnan University, Kunming, China}
\thanks{Z. Zhao is with Institution of Theoretical and Applied Geophysics, School of Earth and Space Sciences, Peking University, Beijing, China, e-mail: zy.zhao@pku.edu.cn (corresponding author)}

\thanks{Manuscript received 6 November 2025; revised 27 March 2026 and 22 June 2026; accepted 10 July 2026. Date of publication 24 July 2026.}
}

\fi

%
%


\markboth{IEEE Transactions on Geoscience and Remote Sensing,~VOL.~64, 2026}{Hu \MakeLowercase{\textit{et al.}}: Bayesian FWI with learned prior using deep CAE}

%



\maketitle

\begin{abstract}
Full waveform inversion (FWI) can be expressed in a Bayesian framework, where the associated uncertainties are captured by the posterior probability distribution (PPD). In practice, solving Bayesian FWI with sampling-based methods such as Markov chain Monte Carlo (MCMC) is computationally demanding because of the extremely high dimensionality of the model space. To alleviate this difficulty, we develop a deep convolutional autoencoder (CAE) that serves as a learned prior for the inversion. The CAE compresses detailed subsurface velocity models into a low-dimensional latent representation, achieving more effective and geologically consistent model reduction than conventional dimension reduction approaches. The inversion procedure employs an adaptive gradient-based MCMC algorithm enhanced by automatic differentiation-based FWI to compute gradients efficiently in the latent space. In addition, we implement a transfer learning strategy through online fine-tuning during inversion, enabling the framework to adapt to velocity structures not represented in the original training set. Numerical experiments with synthetic data show that the method can reconstruct velocity models and assess uncertainty with improved efficiency compared to traditional MCMC methods.

\end{abstract}




%

\IEEEpeerreviewmaketitle

\section{Introduction}
%
%
%
%


\IEEEPARstart{A}{s} a large-scale inverse problem, full waveform inversion (FWI) \cite{lailly1983, tarantola1984} can be formulated within the Bayesian inference framework, yielding {Bayesian FWI} \cite{tarantola1987}. Conventional FWI solves the inverse problem by minimizing the data misfit through local optimization algorithms, without explicitly accounting for model or data uncertainties.  
In contrast, Bayesian FWI incorporates these uncertainties naturally, with the posterior probability distribution (PPD) providing a complete probabilistic description of the solution. We can access statistical assessments, such as mean, media values, variance and covariance of the PPD (e.g. \cite{sen1996}), allowing for more robust and informative inversion results compared to deterministic approaches.

To solve the Bayesian FWI problem, two main strategies are commonly adopted: (i) {variational inference} (VI), which seeks to minimize the Kullback–Leibler (KL) divergence between an approximate distribution and the true PPD \cite{zhang2020, zhang2023}; and (ii) {sampling-based methods}, which draw samples directly from the PPD using Markov chain Monte Carlo (MCMC) methods \cite{martin2012stochastic, visser2019bayesian, guo2020bayesian, zhao2021, Izzatullah2021, hu2023, Berti2024A, hu2025}. While VI offers computational speed through approximation, it sacrifices accuracy by constraining the PPD to a chosen parametric form. Sampling-based methods, though computationally more demanding, avoid such approximations and are therefore preferred when we need to obtain more accurate estimation for PPD.

A primary obstacle for sampling-based Bayesian FWI is the curse of dimensionality, the number of samples required to adequately sample the PPD grows rapidly with the model dimension.  
Reducing the effective dimensionality of the problem can greatly alleviate this issue, enabling more efficient MCMC sampling.  
Various model parameterizations have been proposed and can be employed for this purpose, such as nonoscillatory spline interpolation \cite{Datta2016estimating}, Voronoi tessellation \cite{biswas2017}, Gaussian radial basis functions \cite{chen2020stochastic}, and the discrete cosine transform (DCT) \cite{Berti2024A, hu2024dct}.

Deep neural networks (DNNs) have shown strong potential in geophysical inversion, offering both expressive power and compact representations of high-dimensional models. For example, a hybrid machine learning approach for FWI was developed by compressing seismic data into a latent space using a convolutional autoencoder (CAE)\cite{chen2021seismic}. It is demonstrated that physics-guided CAEs can serve as effective regularizers for both acoustic and elastic FWI \cite{dhara2022physics, dhara2023elastic}. Similarly, it is found that reparameterizing models with deep convolutional neural networks naturally introduces regularization into FWI \cite{zhu2022integrating}. Recently, a deep convolutional autoencoder was implemented for Bayesian FWI as a prior, which is coupled with back-propagated gradient calculated from adjoint-state method \cite{hu2024cae}. An unsupervised variational autoencoder (VAE) approach was proposed for uncertainty analysis in full-waveform inversion \cite{elmeliegy2024uncertainty}.

Recently, seismic forward modeling has been implemented as a recurrent neural network (RNN) \cite{richardson2018seismicfullwaveforminversionusing}, enabling the gradient to be computed by automatic differentiation (AD). Traditionally, the gradient in FWI is obtained using the adjoint-state method \cite{plessix2006review}, which has been shown to be equivalent to those computed through AD \cite{richardson2018seismicfullwaveforminversionusing}. Beyond this equivalence, AD-based approaches provide increased flexibility in the design of inversion frameworks, allowing for more versatile objective functions, easier incorporation of regularization strategies, and seamless integration with neural network architectures \cite{Liu2025}.

\begin{figure*}[!t]
\centering
\includegraphics[width=0.7\linewidth]{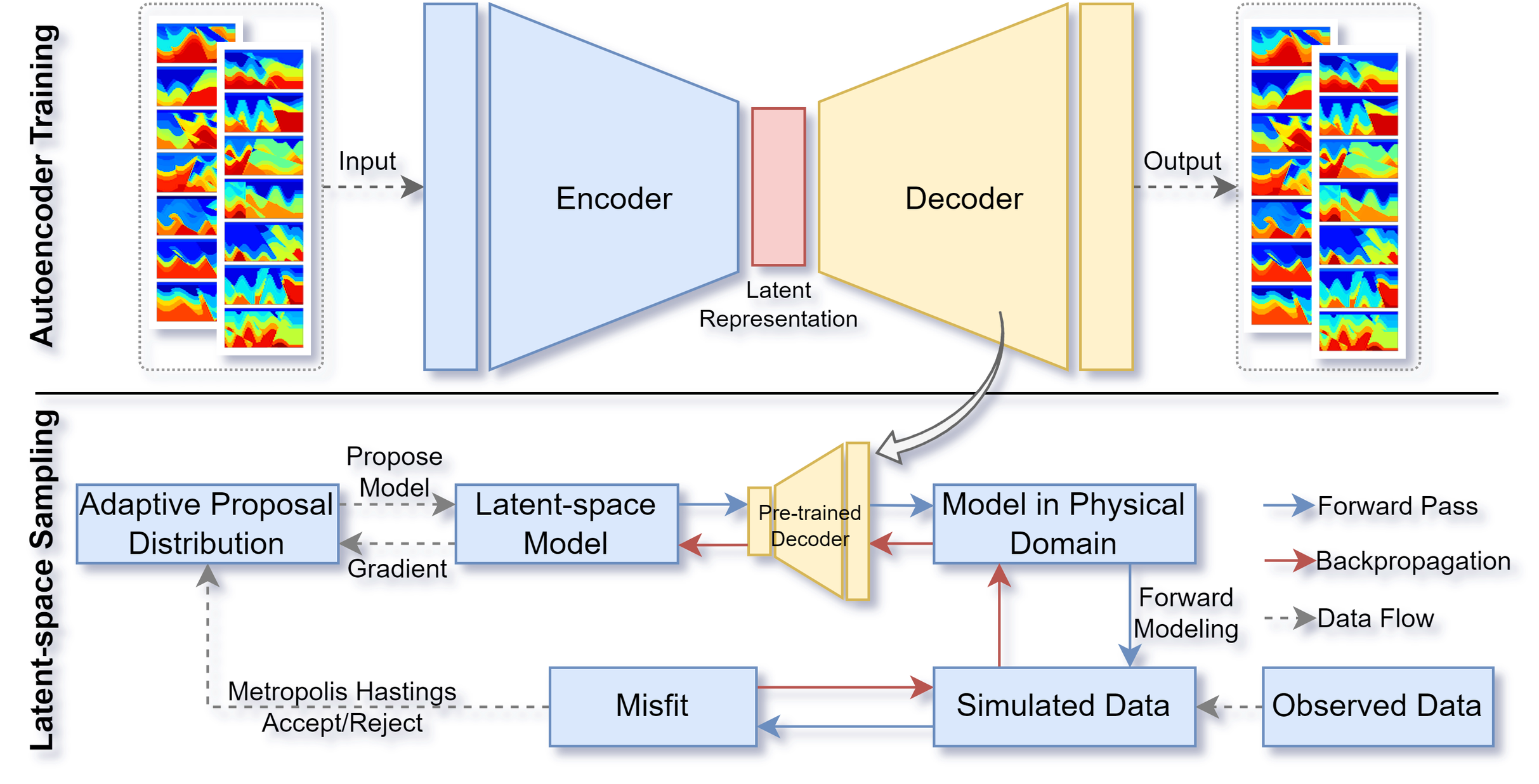}
\caption{The proposed Bayesian FWI approach with learned prior from a pre-trained autoencoder.}
\label{fig:workflow_from_drawio_wAD_Fix_grad.drawio.png}
\end{figure*}

Motivated by these developments, we propose to address the computational challenges of sampling-based Bayesian FWI by training a CAE on a dataset of subsurface models and using its latent representation as a learned prior.  
As illustrated in Figure~\ref{fig:workflow_from_drawio_wAD_Fix_grad.drawio.png}, the CAE maps the high-dimensional model space into a much lower-dimensional latent space. After training, an adaptive gradient-based MCMC sampling process is performed in the latent space, with gradients efficiently computed via automatic differentiation by backpropagating the data misfit to the latent variables. The proposed model in the latent space is transformed to physical domain via the pre-trained decoder, and used for forward modeling. 


Compared to conventional Bayesian FWI approaches, the proposed method significantly improves computational efficiency by compressing the high-dimensional model space into a low-dimensional latent representation, thereby reducing the number of samples required for reliable posterior inference.

In contrast to traditional dimensionality reduction techniques, the data-driven machine learning approach adopted here provides superior model reconstruction capability, enabling more accurate and geologically consistent representations of subsurface structures while facilitating more efficient sampling. 
Furthermore, by incorporating AD-based FWI, gradients can be computed efficiently in the latent space, which not only improves the sampling process but also enables flexible fine-tuning of the learned prior during inversion. 
Finally, the integration of adaptive gradient-based MCMC sampling in the latent space allows for effective exploration of the posterior distribution, mitigating issues such as local minima commonly encountered in deterministic inversion and ensemble collapse often observed in variational inference methods.


\section{METHODS}

\subsection{Bayesian FWI}


Bayesian inference offers a probabilistic framework for solving inverse problems \cite{tarantola1987}. In this approach, the model parameters $\mathbf{m}$ are treated as random variables characterized by probability distributions that quantify their associated uncertainties. According to Bayes’ theorem, the probability distribution of $\mathbf{m}$ conditioned on the observed data $\mathbf{d}$ is given by  
\begin{equation}
P(\mathbf{m}|\mathbf{d}) \propto P(\mathbf{d}|\mathbf{m})\,P(\mathbf{m}),
\label{eq:bayes_rule}
\end{equation}
where the \emph{posteriori} distribution $P(\mathbf{m}|\mathbf{d})$ represents the updated knowledge about $\mathbf{m}$ after incorporating the data. It is proportional to the product of the \emph{prior} distribution $P(\mathbf{m})$, which encodes prior knowledge or assumptions about the model, and the \emph{likelihood} function $P(\mathbf{d}|\mathbf{m})$, which expresses the probability of observing the data $\mathbf{d}$ given the model $\mathbf{m}$.

One can model both the prior distribution and the likelihood function as multivariate Gaussian distributions. Under this assumption, the likelihood can be expressed as\cite{tarantola1987} 
\begin{equation}
P(\mathbf{d}|\mathbf{m}) \propto 
\exp\left\{
    -\frac{1}{2} 
    \left[ \mathbf{d}_\mathrm{obs} - f(\mathbf{m}) \right]^*
    \mathbf{C}_\mathbf{d}^{-1}
    \left[ \mathbf{d}_\mathrm{obs} - f(\mathbf{m}) \right]
\right\},
\label{eq:likelihood}
\end{equation}
where $\mathbf{d}=f(\mathbf{m})$ is a projection from model $\mathbf{m}$ to data $\mathbf{d}$, i.e. forward modeling, $\mathbf{d}_\mathrm{obs}$ is observed data,  $\mathbf{C}_\mathbf{d}$ is the covariance matrix of the observation, and the asterisk represents conjugate transpose. Similarly, the prior distribution can be written as  
\begin{equation}
P(\mathbf{m}) \propto 
\exp\left\{
    -\frac{1}{2} 
    \left( \mathbf{m} - \mathbf{m}_{\mathrm{prior}} \right)^*
    \mathbf{C}_\mathbf{m}^{-1}
    \left( \mathbf{m} - \mathbf{m}_{\mathrm{prior}} \right)
\right\},
\label{eq:prior}
\end{equation}
where $\mathbf{m}_{\mathrm{prior}}$ denotes the prior mean model and $\mathbf{C}_\mathbf{m}$ is the model covariance matrix.

Unlike local optimization methods, which typically yield only a single locally optimal solution, the Bayesian formulation delivers solution of the inverse problem with samples that drawn from the posterior distribution that give a comprehensive quantification of uncertainty in the inversion, encapsulated in the shape and spread of the posterior distribution.

\subsection{Deep convolutional autoencoder}

Autoencoders (AE, \cite{baldi2012autoencoders}) are a class of feedforward neural networks designed to learn an identity mapping \(F\) for a given training dataset, i.e.,
\begin{equation}
    F: \mathbf{m} \mapsto \tilde{\mathbf{m}}, \quad \forall \mathbf{m} \in \mathcal{M},
\label{eq:autoencoder}
\end{equation}
where \(\mathcal{M} = \{\mathbf{m}^{(1)}, \ldots, \mathbf{m}^{(k)}\}\) represents the training set consisting of subsurface models in our context, and \(\tilde{\mathbf{m}} = F(\mathbf{m})\) denotes the reconstructed velocity model produced by the autoencoder.

An AE is composed of two main components: the encoder \(F^{\text{(en)}}\) and the decoder \(F^{\text{(de)}}\). The encoder maps the high-dimensional input \(\mathbf{m}\) into a low-dimensional latent representation \(\bar{\mathbf{m}}\), i.e.,
    $F^{\text{(en)}}: \mathbf{m} \mapsto \bar{\mathbf{m}}$,
while the decoder reconstructs the original input from this latent vector,
    $F^{\text{(de)}}: \bar{\mathbf{m}} \mapsto \tilde{\mathbf{m}}$.
Together, the AE mapping can be expressed as
\begin{equation}
    F: \mathbf{m} \mapsto F^{\text{(de)}} \circ F^{\text{(en)}}(\mathbf{m}), \quad \forall \mathbf{m} \in \mathcal{M},
\end{equation}
where \(\circ\) denotes function composition. Throughout this paper, \(\bar{\mathbf{m}}\) denotes a latent-space velocity model, and \(\tilde{\mathbf{m}}\) denotes the velocity model reconstructed by the AE.

Both encoder and decoder are typically constructed from multiple layers, including input, output, and hidden layers. Assuming both the encoder and decoder has \(l\) layers, the output of each layer is computed by applying a nonlinear transformation to the previous layer’s output:
\begin{equation}
    F^{\text{(en)}}_i: (\mathbf{x}_{i-1}, \bm{\theta}^{\text{(en)}}_i) \mapsto \mathbf{x}_i, \quad i = 1, \ldots, l,
\end{equation}
where \(\bm{\theta}^{\text{(en)}}_i\) represents the trainable parameters (weights and biases) for encoder layer \(i\). The neurons \(\mathbf{x}_i\) are given by
\begin{equation}
    \mathbf{x}_i = \phi\bigl(F^{\text{(en)}}_i(\mathbf{x}_{i-1}, \bm{\theta}^{\text{(en)}}_i)\bigr),
\end{equation}
with \(\phi(\cdot)\) being a nonlinear activation function. Similar equations can be written for the decoder. Denoting \(\Theta = \{\bm{\theta}^{\text{(en)}}_1, \ldots, \bm{\theta}^{\text{(en)}}_L, \bm{\theta}^{\text{(de)}}_1, \ldots, \bm{\theta}^{\text{(de)}}_L \}\) as the collection of all network parameters, the autoencoder training minimizes the following reconstruction loss over the dataset \(\mathcal{M}\):
\begin{equation}
    J(\Theta; \mathcal{M}) = \sum_{i=1}^{k} \left\lVert \mathbf{m}^{(i)} - F^{\text{(de)}} \circ F^{\text{(en)}} \left(\mathbf{m}^{(i)} \right) \right\rVert^2,
\label{eq:autoencoder_loss}
\end{equation}
where \(\|\cdot\|\) denotes the Euclidean (L2) norm.

The layers in both encoder and decoder can be composed of fully connected, convolutional, or hybrid architectures. In this work, we adopt a convolutional autoencoder (CAE) structure motivated by the well-known capability of convolutional neural networks to capture spatially hierarchical features in data \cite{lecun2015deep}. This characteristic makes CAEs particularly suitable for learning compact, abstract latent representation of subsurface models, which often exhibit diverse spatial structures.

The latent dimension \(\bar{\mathbf{m}}\) in a CAE is typically much smaller than the original input dimension, enabling significant dimensionality reduction. Furthermore, the nonlinear activations embedded within the CAE empower the model to capture complex, nonlinear features presented implicitly in the training dataset. Consequently, the training process yields a low-dimensional latent representation that effectively extracts the key characteristics of the subsurface models.

\subsection{Latent space MCMC with AD-based gradient calculation}

In this paper, we employ the adaptive gradient-based MCMC algorithm \cite{hu2023,hu2025} in the latent space to solve the Bayesian FWI problem. The proposal distribution for proposing a new sample can be written as
\begin{equation}
\begin{aligned}
    g(\bar{\mathbf{m}}^\prime|\bar{\mathbf{m}}_i)  = \mathcal{N} \left( \bar{\mathbf{m}}_i + \frac{{\sigma_i}^2}{2} {\bm{\Lambda}}_i {D}(\bar{\mathbf{m}}_i) , {\sigma_i}^2 {\bm{\Lambda}}_i \right), 
\end{aligned}
\label{eq:mala_proposal}
\end{equation}
where $\mathbf{\bar{m}}_i$ and $\mathbf{\bar{m}}^\prime$ are the current and proposed model in latent space respectively. $\sigma_i$ and $\bm{{\Lambda}}_i$ are the adaptive step length and adaptive preconditioning proposal covariance matrix, respectively. Details can be found in \cite{hu2025}. ${D}(\bar{\mathbf{m}}_i)$ is the search direction in latent space, which will be introduced shortly.

The proposed model $\mathbf{\bar{m}}^\prime$ is then transformed to original model domain using the pre-trained decoder, i.e. $\mathbf{\tilde{m}}^\prime = F^{\text{(de)}} (\mathbf{\bar{m}}^\prime) $, and then is accepted with probability of
\begin{equation}
    \begin{aligned}
   \alpha=\min\left(1,\frac{P(\tilde{\mathbf{m}}^\prime | \mathbf{d} )}{P(\tilde{\mathbf{m}}_i | \mathbf{d} )}    \frac{g(\bar{\mathbf{m}}_i|\bar{\mathbf{m}}^\prime)}{g(\bar{\mathbf{m}}^\prime|\bar{\mathbf{m}}_i)} \right),
    \end{aligned}
    \label{eq:acceptprob}
\end{equation}
which is the Metropolis-Hastings (M-H) accept/reject process.

In this paper, we adopt the approach proposed by \cite{richardson2018seismicfullwaveforminversionusing}, in which forward modeling is implemented as a recurrent neural network (RNN). This formulation allows the gradient to be conveniently computed via automatic differentiation (AD).  
Here, we define the search direction in the reconstructed model space as  
\begin{equation}
    D(\tilde{\mathbf{m}}_i) = \nabla \log P(\mathbf{d} \,|\, \tilde{\mathbf{m}}_i),
\end{equation}
which corresponds to the derivative of the log-likelihood and can be computed by backpropagation (BP; \cite{Werbos1990}) of the data misfit through the RNN.  

Using the chain rule, the search direction in the latent space, $D(\bar{\mathbf{m}}_i)$, can be expressed as  
\begin{equation}
    D(\bar{\mathbf{m}}_i) 
    = \nabla \log P(\mathbf{d} \,|\, \tilde{\mathbf{m}}_i) 
    \cdot \frac{\partial \tilde{\mathbf{m}}}{\partial \bar{\mathbf{m}}},
\label{eq:search_direction}
\end{equation}
where $\tilde{\mathbf{m}}$ and $\bar{\mathbf{m}}$ are connected through the decoder,
\begin{equation}
   F^{\text{(de)}}: (\bar{\mathbf{m}}, \bm{\theta}^{\text{(de)}}) \rightarrow \tilde{\mathbf{m}}. 
\label{eq:decoder}
\end{equation}
Thus, the latent-space search direction $D(\bar{\mathbf{m}}_i)$ in equation~(\ref{eq:mala_proposal}) can be obtained via a single backpropagation of the data loss through the RNN, followed by the decoder, using equation (\ref{eq:search_direction}).

Traditionally, the gradient is computed using the adjoint-state method \cite{plessix2006review}. It has been proven that gradients obtained via the adjoint-state method are equivalent to those computed with AD \cite{richardson2018seismicfullwaveforminversionusing}. However, compared to traditional approach that backpropagates the adjoint-state gradient through the network (e.g. \cite{hu2024cae}), the AD-based methods offers greater flexibility in designing objective functions, applying regularization techniques, and integrating with neural networks \cite{Liu2025}. In our implementation, the gradient of the data loss with respect to the decoder parameters, as shown in equation~(\ref{eq:decoder}), can also be computed. This enables a flexible fine-tuning strategy, which will be discussed later.

\begin{algorithm}
\caption{AD-based adaptive MCMC sampling in a learned latent space}
\begin{algorithmic}[1]
    \STATE Initialize step length $\sigma_0$ and preconditioning matrix ${\bm{\Lambda}}_0$.
    \STATE Sample an initial model $\mathbf{m}_0 \sim P(\mathbf{m})$.
    \STATE Encode to latent space: $\mathbf{\bar{m}}_0 = F^{\text{(en)}}(\mathbf{m}_0)$.
    \STATE Decode to physical space: $\tilde{\mathbf{m}}_0 = F^{\text{(de)}}(\mathbf{\bar{m}}_0)$.
    \STATE Compute AD-based search direction ${D}(\mathbf{\bar{m}}_{0})$ using Eq.~\eqref{eq:search_direction}.
    \FOR{$i = 0$ to $N_{\text{burnin}} + N_{\text{draws}}$}
        \STATE Generate proposal $\mathbf{\bar{m}}^\prime$ from Eq.~\eqref{eq:mala_proposal}.
        \STATE Forward pass to compute FWI misfit and posterior probability using Eq.~\eqref{eq:bayes_rule}.
        \STATE Compute acceptance probability $\alpha$ using Eq.~\eqref{eq:acceptprob}.
        \STATE Draw $u \sim \mathcal{U}(0,1)$.
        \IF{$u < \alpha$}
            \STATE Accept proposal: $\mathbf{\bar{m}}_{i+1} = \mathbf{\bar{m}}^\prime$.
            \STATE Compute AD-based search direction ${D}(\mathbf{\bar{m}}_{i+1})$ using Eq.~\eqref{eq:search_direction}.
        \ELSE
            \STATE Reject proposal: $\mathbf{\bar{m}}_{i+1} = \mathbf{\bar{m}}_{i}$.
        \ENDIF
        \STATE Update adaptive step length $\sigma_{i+1}$.
        \STATE Update adaptive preconditioning matrix $\bm{{\Lambda}}_{i+1}$.
    \ENDFOR
\end{algorithmic}
\label{algo:dct_mcmc_fwi_pt_algorithm}
\end{algorithm}





The proposed method, outlined in Algorithm~\ref{algo:dct_mcmc_fwi_pt_algorithm}, performs adaptive MCMC sampling in a learned latent space, leveraging AD-based gradient calculation to guide proposals. We first initialize the MCMC parameters, including the step size $\sigma_0$ and preconditioning matrix $\bm{\Lambda}_0$. An initial model $\mathbf{m}_0$ is drawn from the prior distribution $P(\mathbf{m})$ and projected into the latent space through the encoder $F^{(\mathrm{en})}$. The corresponding physical model is reconstructed by the decoder $F^{(\mathrm{de})}$, enabling evaluation of the forward modeling and data misfit. 
The search direction $D(\bar{\mathbf{m}})$ in the latent space is obtained using AD as defined in equation~(\ref{eq:search_direction}).

At each iteration, a proposal $\bar{\mathbf{m}}'$ is generated using a Langevin-type update (equation~\ref{eq:mala_proposal}). This proposal is decoded, and the corresponding posterior probability is computed via Bayes' rule (equation~\ref{eq:bayes_rule}). The Metropolis–Hastings acceptance criterion is applied using equation~(\ref{eq:acceptprob}). If accepted, the latent model is updated and the new search direction is recomputed using AD; otherwise, the state is retained. Both the step size and preconditioning matrix are adaptively tuned during the sampling process, improving convergence and mixing efficiency. This framework significantly reduces the dimensionality of the sampling space while preserving gradient information through AD, thus enhancing sampling efficiency compared to direct sampling in the original high-dimensional physical model space.

\subsection{Transfer learning through online fine-tuning}

Given a training dataset of subsurface velocity models, the autoencoder is designed to learn a compact, low-dimensional representation of the model features in a latent space. During Bayesian inversion, the algorithm can effectively recover a latent representation that matches the observed seismic data, provided that the target model shares similar features to those of the training set. We refer to this scenario as in-distribution (ID) inversion. In contrast, when the target model contains features that deviate significantly from the training distribution, known as the out-of-distribution (OOD) case, the inversion performance typically degrades.

\begin{figure}
    \centering
    \includegraphics[width=0.8\linewidth]{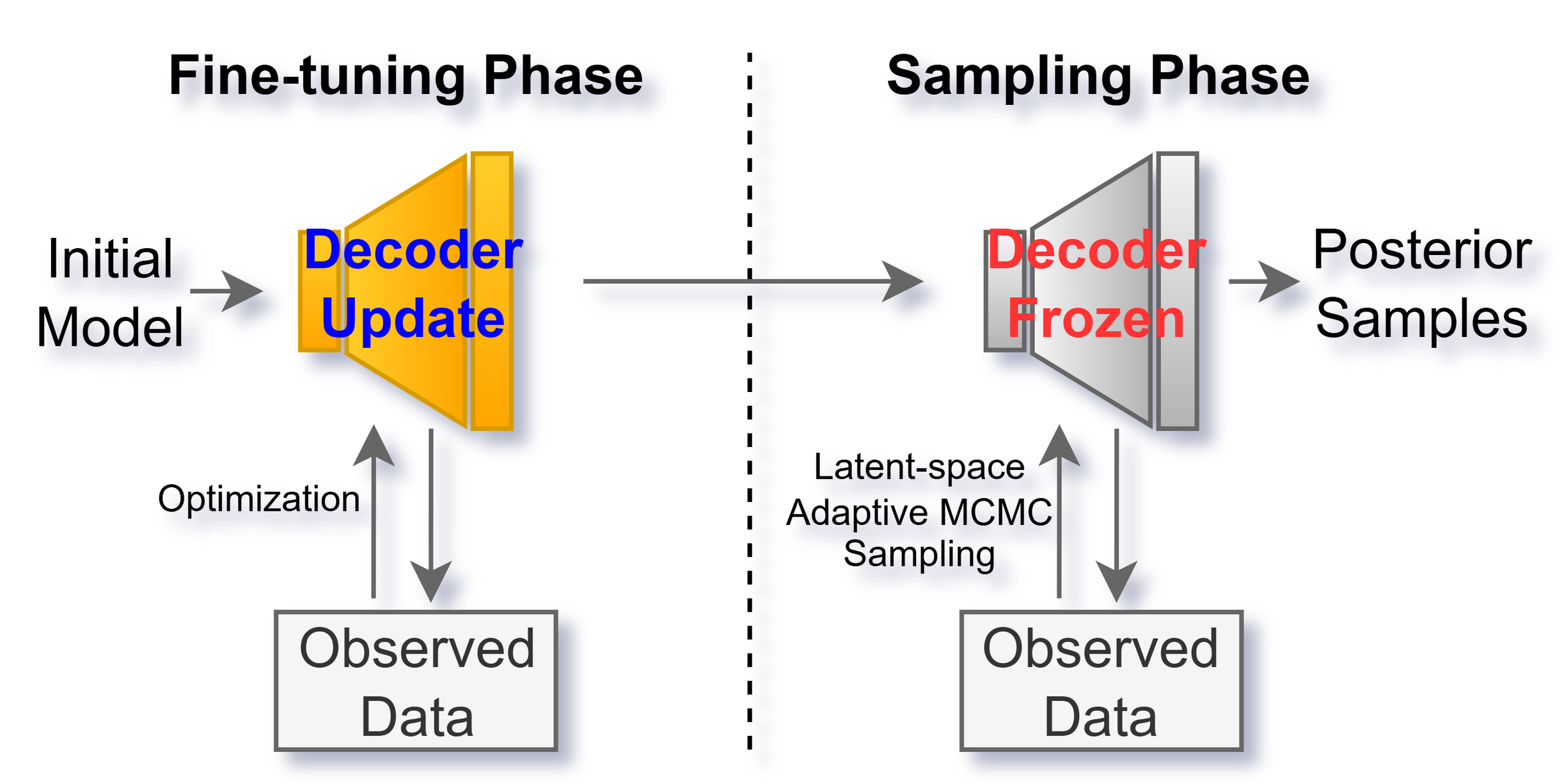}
    \caption{Schematic plot showing the online fine-tuning process during inversion.}
    \label{fig:fine_tune/online_finetuning_plot.drawio.png}
\end{figure}

To address the limitations of out-of-distribution inversion, we propose a transfer learning strategy based on online fine-tuning. A schematic overview of the proposed approach is illustrated in Figure~\ref{fig:fine_tune/online_finetuning_plot.drawio.png}. After pre-training the autoencoder, we retain the decoder for use in the inversion process, which consists of two distinct phases: a fine-tuning phase and a sampling phase. The fine-tuning phase occurs at the early stage of inversion and involves only a few tens of iterations. This phase can be viewed as an optimization step, where the misfit between the synthetic and observed seismic data is used to compute gradients with respect to the decoder weights. These weights are then updated using a gradient-based optimization algorithm, such as Adam. Once fine-tuning is complete, the decoder weights are fixed, and the inversion proceeds to the latent-space sampling phase. In this phase, we employ the adaptive MCMC \cite{hu2025} algorithm to draw posterior samples from the latent space.

\section{EXAMPLE}

\subsection{Dataset and inversion scheme}

\begin{figure*}[!t]
    \centering
    \includegraphics[width=0.9\linewidth]{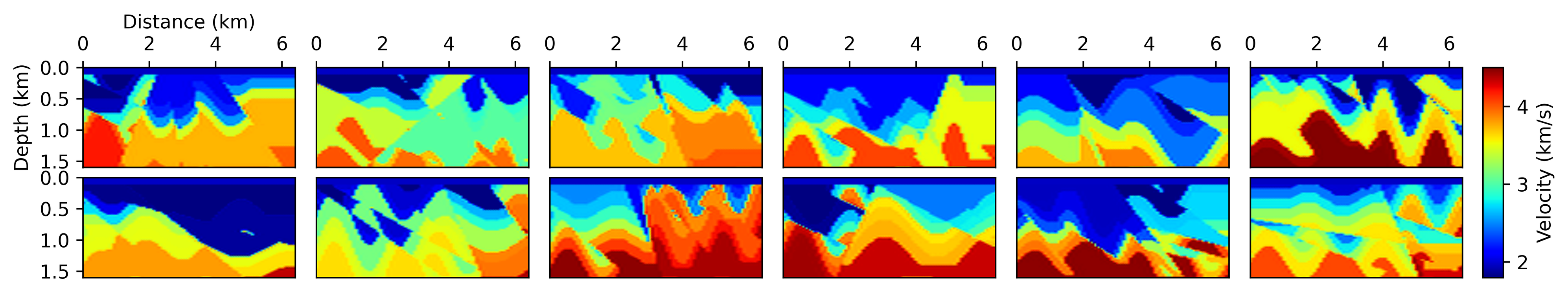}
    \caption{Typical velocity models of the dataset used for training the CAE.}
    \label{fig:2025-05-29_openFWI_CurveFaultB_model_plot_new.png}
\end{figure*}

The quality and diversity of the training dataset are critical to the success of autoencoder pre-training. In this study, we employ the OpenFWI dataset \cite{deng2022openfwi}, a well-established synthetic benchmark, to evaluate our proposed method. This dataset includes a wide variety of geological scenarios, such as curved horizons and complex fault structures. Representative velocity models from the subset we use are shown in Figure~\ref{fig:2025-05-29_openFWI_CurveFaultB_model_plot_new.png}. To enhance subsurface illumination, we modify the original model grid from 70 × 70 to 64 × 128. Out of the 54,000 2D velocity models, we allocate 80\% for training and the remaining 20\% for validation.


For the Bayesian inversion, we set the model grid spacing to \( d_x = d_z = 25\,\text{m} \). As a result, there are 8192 model parameters. The acquisition geometry consists of 32 shots with a shot spacing of 100\,m, and 128 receivers spaced every 25\,m. Both sources and receivers are placed within the shallow water layer. A Ricker wavelet with an 8\,Hz peak frequency is used as the source wavelet for both forward modeling and inversion. To simulate measurement noise, we add zero-mean additive white Gaussian noise (AWGN) to the synthetic data, such that the signal-to-noise ratio (SNR) is 10\,dB. Specifically, the noisy data \( \tilde{\mathbf{d}} \) is given by
\[
\tilde{\mathbf{d}} = \mathbf{d} + \Psi, \quad \Psi \sim \mathcal{N}(0, \sigma_{\mathbf{d}}^2 \mathbf{I}),
\]
where \( \mathbf{d} \) is the clean synthetic data, and \( \sigma_{\mathbf{d}} \) is chosen to achieve the desired SNR. All inversion examples run for the same 5,000 iterations on a single NVIDIA RTX A5000 GPU. The wall-clock running time for all inversion numerical examples in this paper are presented in Table \ref{tab:fwi_running_time}. Note that the same grid size is used for different frequency inversions, resulting a similar running time for low frequency inversion (out-of-distribution inversion) and high-frequency inversion (extension to higher frequency).

\newcolumntype{Y}{>{\raggedright\arraybackslash}X}

\begin{table}[htbp]
    \centering
    \caption{Wall-clock running time for all inversion examples presented in this paper.}
    \label{tab:fwi_running_time}
    \begin{tabularx}{\columnwidth}{Yll}
        \toprule
        Numerical Example & Model Parameterization & Running Time \\
        \midrule
        \multirow{3}{*}{\parbox{2.8cm}{In-distribution inversion \\ (Figure \ref{fig:in_dist_inversion/in_distribution_CAE_MCMC_inversion_fwi.png})} }
            & No compression & 18 min 55 s \\
            & DCT            & 24 min 35 s \\
            & CAE            & 24 min 32 s \\
        \midrule
        \multirow{4}{*}{\parbox{2.8cm}{Out-of-distribution inversion (Figure \ref{fig:fine_tune/2025-08-05-09-18-39CAE_MCMC_inversion_example_fwi.png})} }
            & No compression       & 19 min 7 s \\
            & DCT                  & 24 min 31 s \\
            & CAE                  & 23 min 51 s \\
            & CAE with fine-tuning  & 24 min 20 s \\
        \midrule
        \multirow{3}{*}{\parbox{2.8cm}{Extension to higher frequency (Figure \ref{fig:rev2/MCMC_inversion_example_fwi_all4comparision_paper.png})} }
            & No compression       & 18 min 45 s \\
            & DCT                  & 25 min 5 s \\
            & CAE with fine-tuning  & 25 min 19 s \\        
        \bottomrule
    \end{tabularx}
\end{table}

\subsection{CAE structure and pre-training}

\begin{figure*}[!t]
    \centering
    \includegraphics[width=0.9\linewidth]{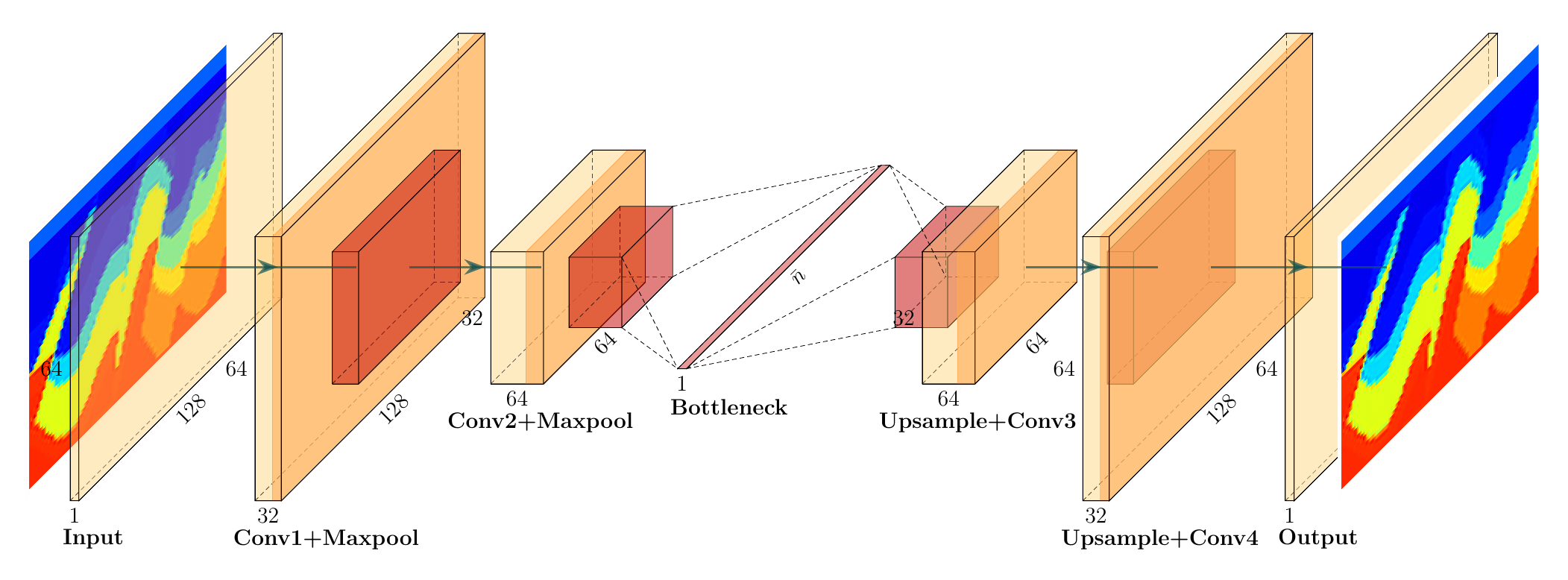}
    \caption{The structure of the convolutional autoencoder used in this paper.}
    \label{fig:CAE_structure_plot.pdf}
\end{figure*}

The structure of the convolutional autoencoder used in this study is illustrated in Figure~\ref{fig:CAE_structure_plot.pdf}. The encoder consists of two convolutional layers with 32 and 64 channels, respectively. Each layer uses a convolutional kernel of size \(3 \times 3\). To reduce the dimensionality of the extracted features, each convolutional layer is followed by a max-pooling layer with a kernel size of \(2 \times 2\).

Deconvolution operations in neural networks are known to produce checkerboard artifacts \cite{odena2016deconvolution}. To mitigate this issue, we adopt an upsampling-convolution strategy in the decoder. Specifically, upsampling is performed via nearest-neighbor interpolation, followed by a convolutional layer with a \(3 \times 3\) kernel. Similar to the encoder, the decoder consists of two convolutional layers with 64 and 32 channels, respectively. In our architecture, convolutional layers are used purely for feature extraction, without altering the spatial resolution. Downsampling is performed using max-pooling layers, while upsampling is handled by interpolation layers.

The encoder and decoder are connected through a flatten-reshape operations and a fully connected bottleneck layer denoted as \( \bar{\mathbf{m}} \), with latent dimension \( \bar{n} \). The latent dimension is significantly smaller than the input dimension \( n_x \times n_z \), enabling the network to learn a compact representation. To investigate the effect of latent space size, we experiment with a ladder of values \( \bar{n}_j = 2^j \) for \( j = 3, \ldots, 7 \). Details of this study and corresponding training results are provided in Appendix~\ref{sec:appendix_latent_dim}.

Unlike the commonly used rectified linear unit (ReLU) activation function in convolutional neural networks \cite{nair2010rectified}, we employ a sine activation function for each convolutional layer. Our experiments show that while ReLU performs slightly better during autoencoder pre-training with relatively small latent dimensions, the sine activation function is more effective in high-latent-dimension settings and in the fine tuning process for the transfer learning task. Additional comparisons and results are provided in Appendix~\ref{sec:appendix_activation}.


\begin{figure*}[!t]
    \centering
    \includegraphics[width=0.7\linewidth]{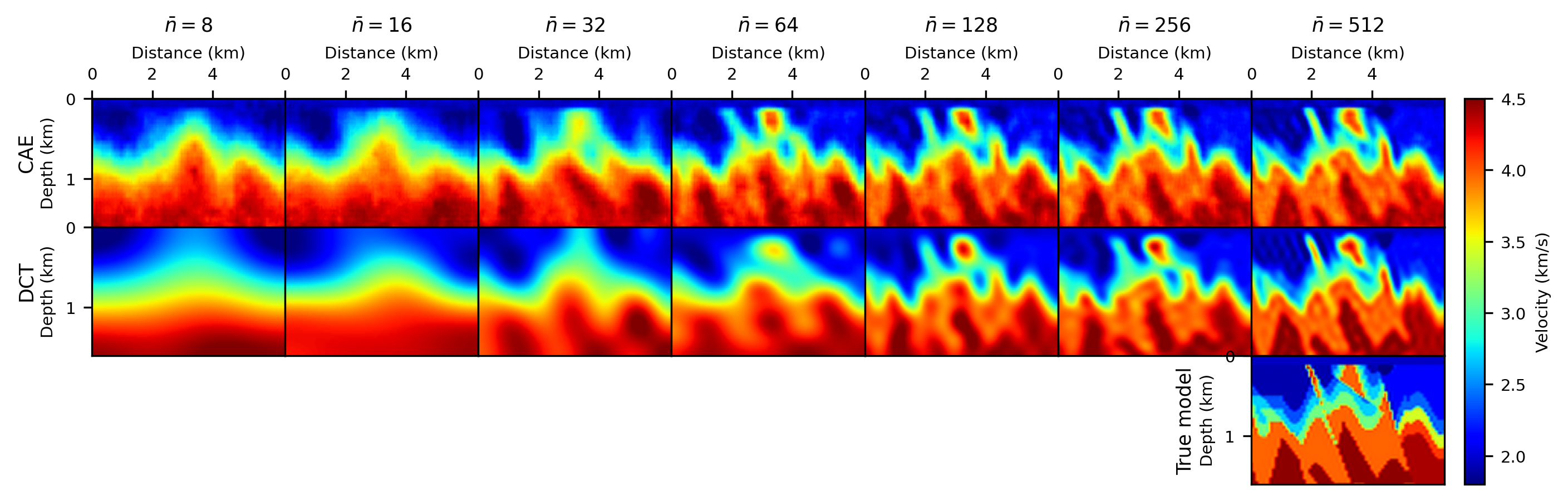}
    \caption{After training of the CAE, we verify the trained autoencoder by implementing model compression on one sample from the test dataset. Also we compare the performance on dimensionality reduction of CAE with discrete cosine transform (DCT).}
    \label{fig:2025-05-30__upsample_nearest_sine_DCT_comparison_vs_CAE_comparison_8_16_32_64_128.png}
\end{figure*}

\begin{figure}[!t]
    \centering
    \includegraphics[width=0.7\linewidth]{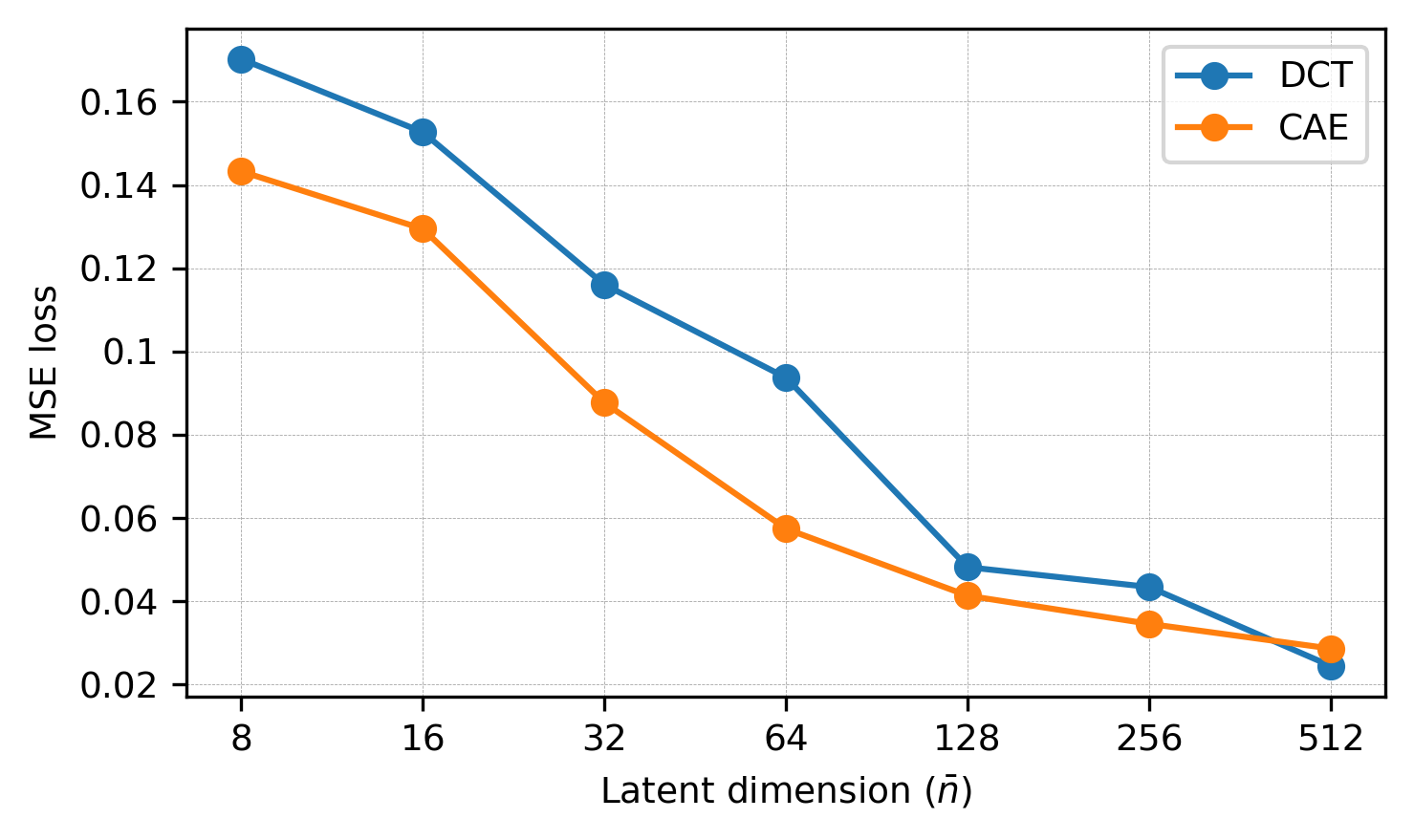}
    \caption{The MSE loss comparison of CAE (blue) and DCT (orange) for different latent dimensions as shown in Figure \ref{fig:2025-05-30__upsample_nearest_sine_DCT_comparison_vs_CAE_comparison_8_16_32_64_128.png}.}
    \label{fig:2025-05-30__upsample_nearest_sine_DCT_comparison_vs_CAE_comparison_MSE_loss_curve_plot.png}
\end{figure}

After training, we validate the performance of the CAE using the test dataset and compare its effectiveness with a classical dimensionality reduction technique: the discrete cosine transform (DCT). Figure~\ref{fig:2025-05-30__upsample_nearest_sine_DCT_comparison_vs_CAE_comparison_8_16_32_64_128.png} presents one representative velocity model from the test set, along with reconstructions obtained from both CAE and DCT at different low dimensions. Both methods successfully preserve the low-wavenumber components of the velocity model, though the CAE consistently outperforms DCT for very low dimensions. Figure~\ref{fig:2025-05-30__upsample_nearest_sine_DCT_comparison_vs_CAE_comparison_MSE_loss_curve_plot.png} displays the mean squared error (MSE) associated with the reconstructions shown in Figure~\ref{fig:2025-05-30__upsample_nearest_sine_DCT_comparison_vs_CAE_comparison_8_16_32_64_128.png}. Except for the case when \( \bar{n} = 512 \), the CAE yields significantly lower MSE compared to DCT, indicating better reconstruction accuracy.

\begin{figure}[!t]
    \centering
    \includegraphics[width=\linewidth]{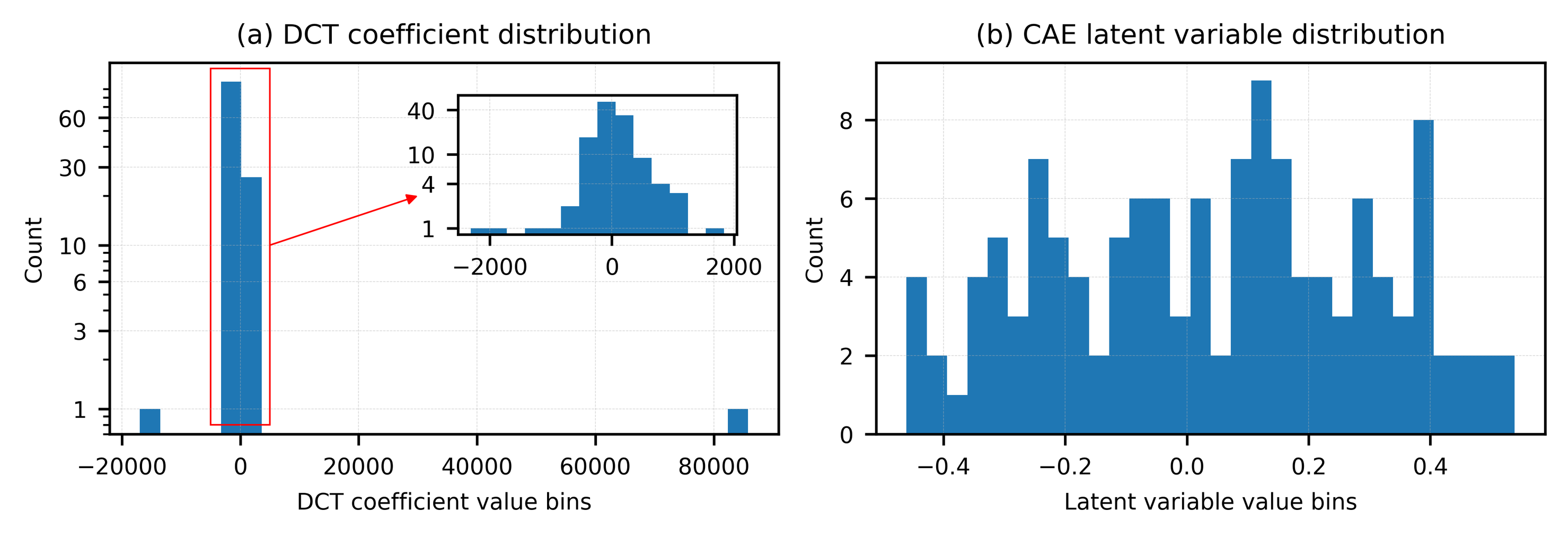}
    \caption{Comparison of latent space variable value distribution for DCT (a) and CAE (b), both for the same model as shown in Figure \ref{fig:2025-05-30__upsample_nearest_sine_DCT_comparison_vs_CAE_comparison_8_16_32_64_128.png}, and 128 latent variables.}
    \label{fig:2025-05-30__upsample_nearest_sine_DCT_comparison_vs_CAE_comparison_latent_dist.png}
\end{figure}

In addition, we analyze the distribution of latent variables for the same test model, as shown in the histogram plot in Figure~\ref{fig:2025-05-30__upsample_nearest_sine_DCT_comparison_vs_CAE_comparison_latent_dist.png}. Note that the y-axis scales are different between the two histograms to highlight the presence of outliers in the DCT coefficients. The DCT coefficients as a latent representation are heavily concentrated near zero, but with a small number of extreme values extending far into both positive and negative directions. The inset plot reveals a roughly symmetric, bell-shaped core, but the overall distribution exhibits heavy tails and strong deviations from normality. This presents challenges for sampling algorithms, as significant differences in the scaling of latent coefficients must be accounted for. For example, \cite{hu2024dct} proposed a modified version of the Yeo–Johnson transform to address this issue. In contrast, the CAE latent variables exhibit a well-scaled distribution with no extreme outliers, which is more favorable for MCMC sampling and facilitates a more stable and efficient inference.

\subsection{In-distribution inversion}

\begin{figure*}[!t]
    \centering
    \includegraphics[width=0.8\linewidth]{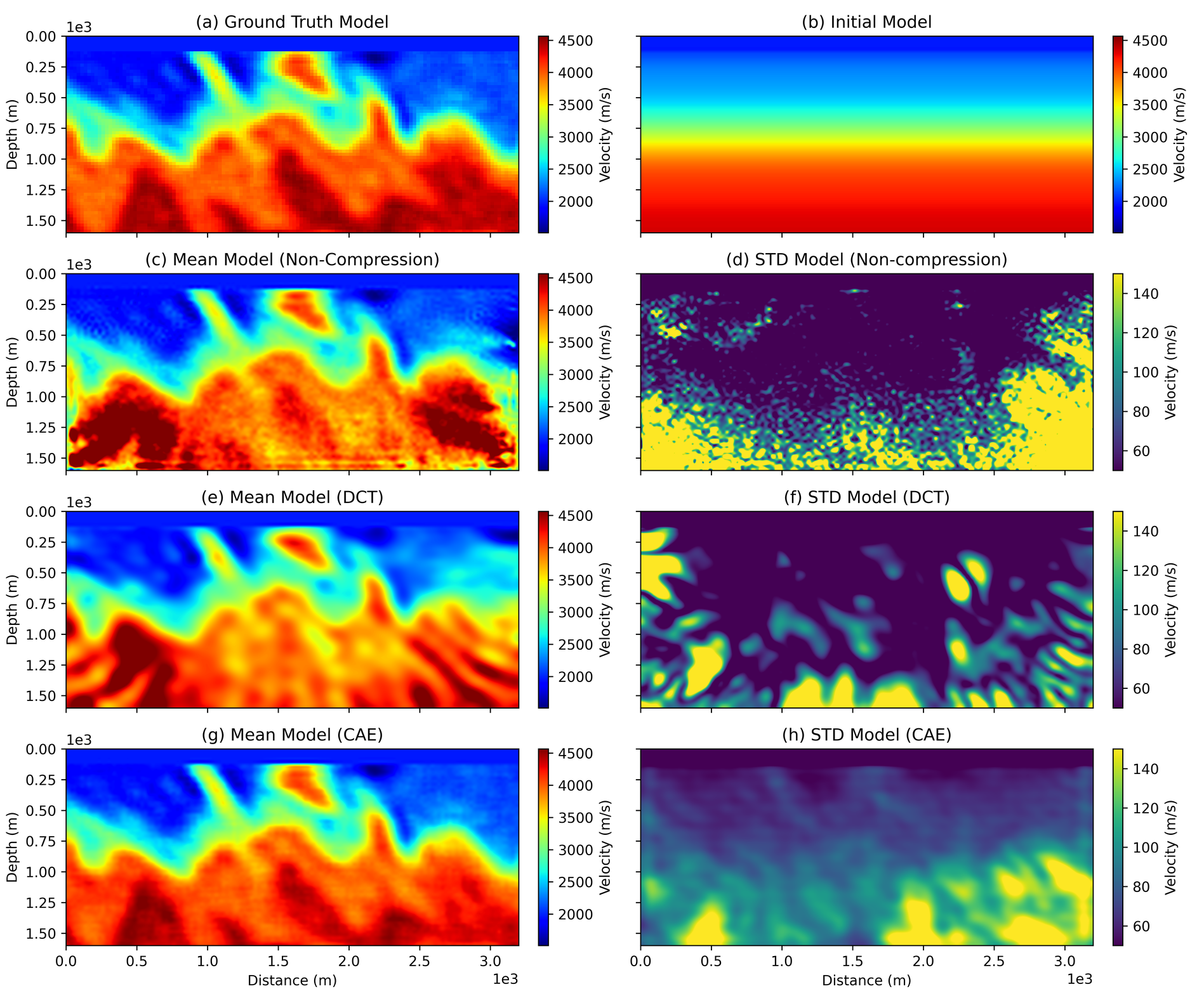}
    \caption{We use the CAE compressed model as true velocity model (a). Starting from a 1D gradient velocity model (b), the proposed method produce the inverted mean model (g) and quantified uncertainties (h) with 5,000 iterations. As a comparison, we show the inverted mean model (c) and standard deviation model (d) without using dimensionality reduction, as well as DCT-based dimensionality reduction (e \& f).}
    \label{fig:in_dist_inversion/in_distribution_CAE_MCMC_inversion_fwi.png}
\end{figure*}

\begin{figure*}[!t]
    \centering
    \includegraphics[width=0.8\linewidth]{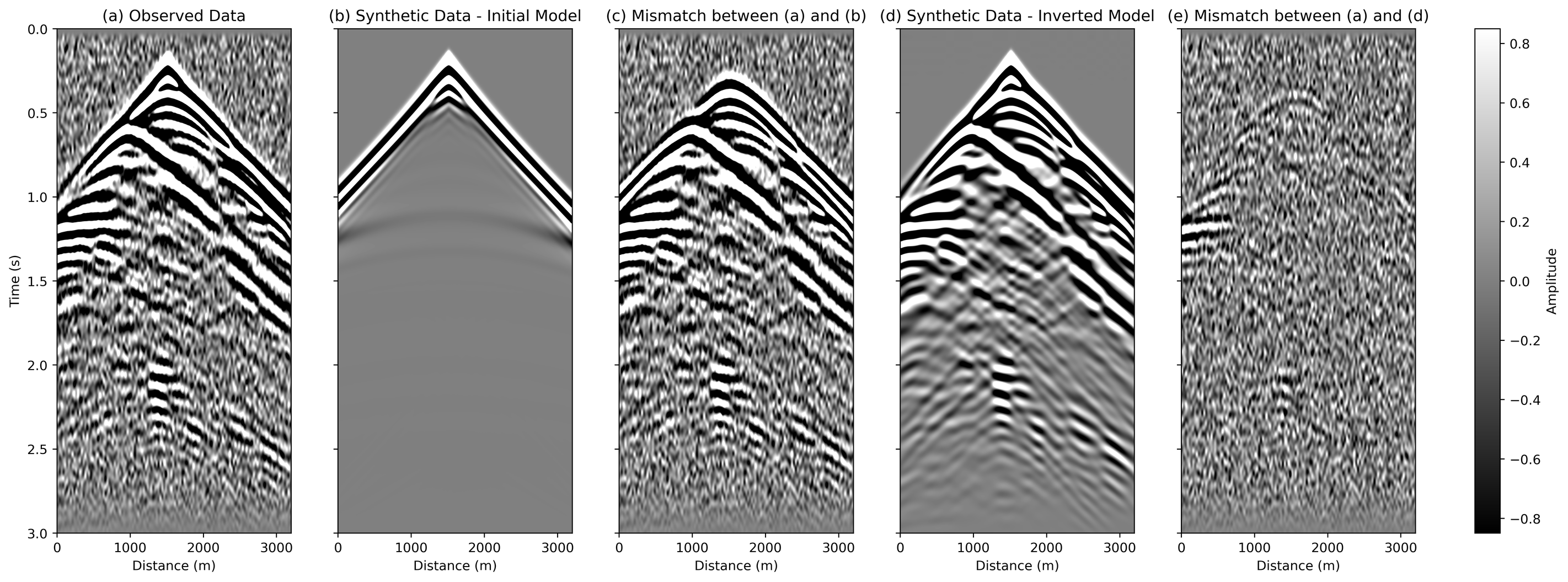}
    \caption{We compare the observed shot gather (a) with the synthetic shot gather from initial model (b) with mismatch (c). After inversion, the synthetic shot gather using inverted model (d) match the observed data very well, with mismatch shown in (e).} 
    \label{fig:in_dist_inversion/shot_gather_comparison_observed_synthetic_inverted.png}
\end{figure*}

To verify the effectiveness of the proposed method for in-distribution (ID) inversion, we select one compressed velocity model from the test dataset (Figure~\ref{fig:in_dist_inversion/in_distribution_CAE_MCMC_inversion_fwi.png}a) and use it as the ground-truth model. The observed data are generated using this true model and the acquisition parameters described in the previous section. For inversion, we use the observed data with frequency up to 15 Hz. An example shot gather from the observed data is shown in Figure~\ref{fig:in_dist_inversion/shot_gather_comparison_observed_synthetic_inverted.png}a. The inversion starts from a simple 1D initial model (Figure~\ref{fig:in_dist_inversion/in_distribution_CAE_MCMC_inversion_fwi.png}b). The corresponding synthetic shot gather is shown in Figure~\ref{fig:in_dist_inversion/shot_gather_comparison_observed_synthetic_inverted.png}b, and the data mismatch is displayed in Figure~\ref{fig:in_dist_inversion/shot_gather_comparison_observed_synthetic_inverted.png}c.

\begin{figure}[!t]
    \centering
    \includegraphics[width=\linewidth]{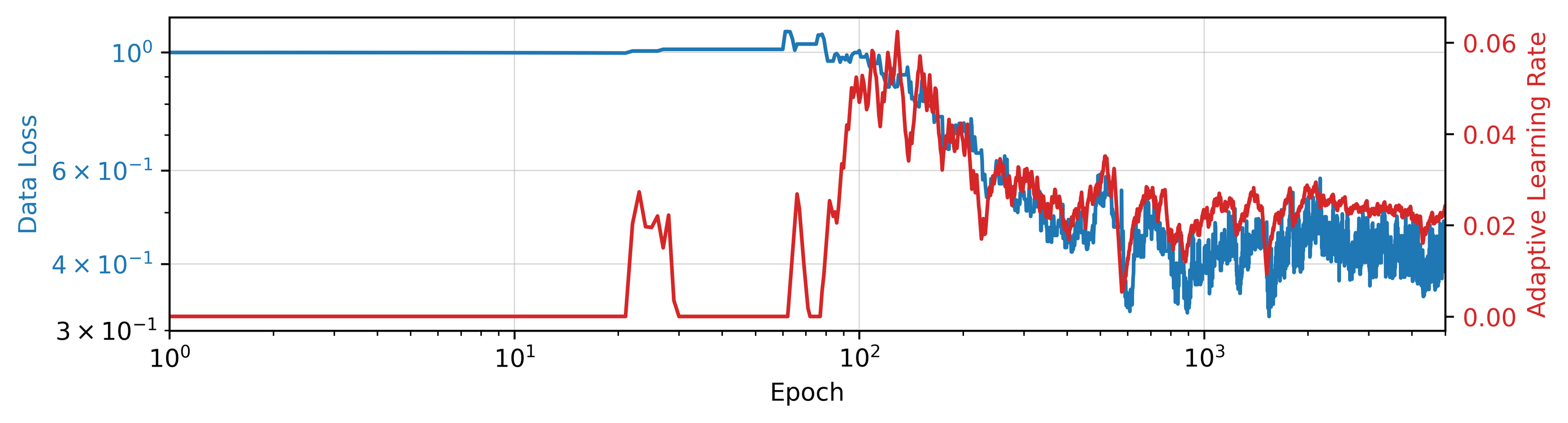}
    \caption{The evolution of FWI misfit (blue) and adaptive step length (red) versus MCMC iterations.}
    \label{fig:in_dist_inversion/in_distribution_data_loss_plot_with_adaptive_lr.png}
\end{figure}

The inversion runs for 5000 iterations. Figure~\ref{fig:in_dist_inversion/in_distribution_data_loss_plot_with_adaptive_lr.png} illustrates the evolution of the data misfit and the adaptive learning rate during the inversion. The adaptation mechanism begins after 20 iterations, allowing the algorithm to automatically adjust the step size for optimal sampling efficiency. The burn-in period is set to \( N_{\text{burn-in}} = 500 \), and all samples beyond this phase are used for posterior statistical estimation.

After inversion, the mean model (Figure~\ref{fig:in_dist_inversion/in_distribution_CAE_MCMC_inversion_fwi.png}g) is calculated by arithmetic mean of all posterior samples, and treated as the inverted model. It closely matches the true velocity model. This agreement is supported by the comparison between the observed and synthetic shot gather. The synthetic gather generated from the inverted model is shown in Figure~\ref{fig:in_dist_inversion/shot_gather_comparison_observed_synthetic_inverted.png}d, while their difference is displayed in Figure~\ref{fig:in_dist_inversion/shot_gather_comparison_observed_synthetic_inverted.png}e. The residual primarily consists of random noise, indicating that the inverted model accurately reproduces the observed data.

More importantly, our method provides a posterior estimate of uncertainty through the ensemble of posterior samples. As shown in Figure~\ref{fig:in_dist_inversion/in_distribution_CAE_MCMC_inversion_fwi.png}h, regions with poor illumination — particularly in the lower corners — exhibit higher uncertainty, highlighting the contributing factor of acquisition geometry on inversion uncertainty.

For comparison, we also present inversion results obtained using (i) adaptive MCMC sampling without dimensionality reduction (Figure~\ref{fig:in_dist_inversion/in_distribution_CAE_MCMC_inversion_fwi.png}c--d), and (ii) a DCT-based model dimensionality reduction approach (Figure~\ref{fig:in_dist_inversion/in_distribution_CAE_MCMC_inversion_fwi.png}e--f). 
For a fair comparison, all methods are applied to the same synthetic dataset, initialized from the same starting model, use the same adaptive gradient-based MCMC algorithm, and are run for 5,000 iterations.

Without dimensionality reduction, the inversion is only able to recover the shallow portion of the velocity model (Figure~\ref{fig:in_dist_inversion/in_distribution_CAE_MCMC_inversion_fwi.png}c). 
This limitation arises from the high dimensionality of the model space, where a significantly larger number of samples is typically required to achieve adequate exploration of the posterior distribution. 
Although the inverted standard deviation model (Figure~\ref{fig:in_dist_inversion/in_distribution_CAE_MCMC_inversion_fwi.png}d) reflects an uncertainty pattern consistent with the acquisition geometry, it is heavily contaminated by noise. 
This noise originates from the fine-scale Gaussian perturbations introduced during the sampling process, indicating insufficient convergence of the Markov chain.

For the DCT-based approach, the latent dimensionality is chosen to match that of the CAE for a fair comparison. 
The inverted mean model (Figure~\ref{fig:in_dist_inversion/in_distribution_CAE_MCMC_inversion_fwi.png}e) successfully reconstructs most of the large-scale features of the true model; however, it is significantly affected by DCT-induced artifacts. 
In particular, diagonal striping patterns are observed in poorly illuminated regions. 
Similarly, the corresponding STD model (Figure~\ref{fig:in_dist_inversion/in_distribution_CAE_MCMC_inversion_fwi.png}f) exhibits pronounced artifacts, leading to biased and spatially inconsistent uncertainty estimates.

Among the three approaches, the CAE-based method produces the most accurate reconstruction of the velocity model and provides more reliable uncertainty quantification within a limited number of iterations. 
This result highlights the advantage of data-driven dimensionality reduction in improving both inversion quality and sampling efficiency for Bayesian FWI.

\begin{figure*}[!t]
    \centering
    \includegraphics[width=0.7\linewidth]{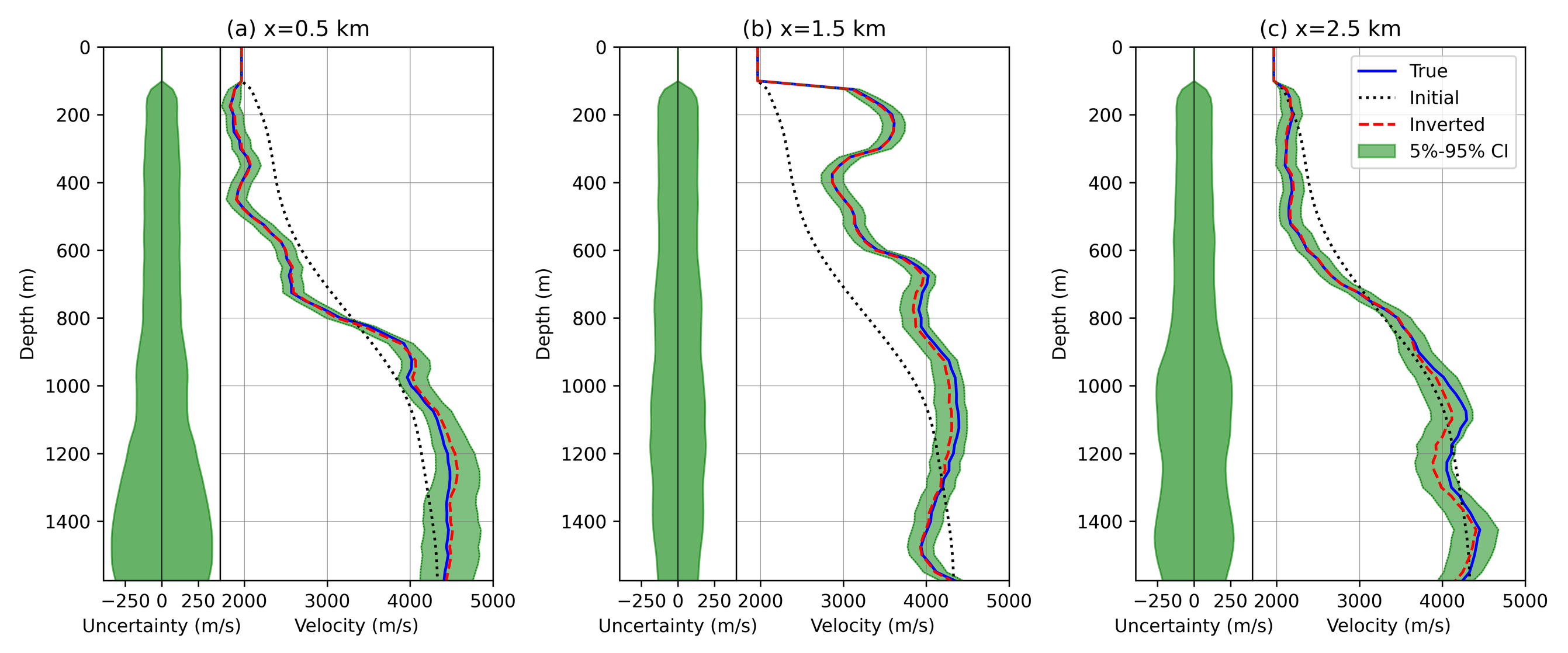}
    \caption{Vertical trace plot of the inversion results as shown in Figure \ref{fig:in_dist_inversion/in_distribution_CAE_MCMC_inversion_fwi.png} at three horizontal locations, 0.5km (a), 1.5km (b) and 2.5km (c). We show the true model (blue), initial model (grey dotted), inverted model (red dashed), as well as the 5\% to 95\% confidence interval (blue shaded). }
    \label{fig:in_dist_inversion/in_distribution_CAE_MCMC_inversion_traceplot.png}
\end{figure*}

Finally, we analyze trace plots of the inversion results at three different horizontal locations, presented in Figure~\ref{fig:in_dist_inversion/in_distribution_CAE_MCMC_inversion_traceplot.png}. The red dashed line indicates the inverted mean model, while the blue line denotes the true model. The conformation between both lines indicates a low reconstruction error. The confidence interval (CI) between 5th and 95th percentiles is represented as the green shaded area. As depth increases, the confidence interval widens, reflecting greater uncertainty in the deeper sections of the model.





\subsection{Out-of-distribution direct inversion}



It is of particular interest to evaluate the performance of the pre-trained CAE when applied to an unseen dataset that is not represented in the training dataset. In this section, we present an example of direct Bayesian inversion using the pre-trained CAE on an out-of-distribution dataset. 

The synthetic observed data are generated from the Marmousi model, as shown in Figure~\ref{fig:fine_tune/2025-08-05-09-18-39CAE_MCMC_inversion_example_fwi.png}a, using the same acquisition geometry described in the previous section. For the inversion, we limit the frequency content to 5\,Hz in order to emphasize the recovery of low-wavenumber structures. An example shot gather from the observed dataset is shown in Figure~\ref{fig:fine_tune/2025-08-05-09-18-39_observed_simulated_data.png}a. The inversion is initialized from a 1D velocity model (Figure~\ref{fig:fine_tune/2025-08-05-09-18-39CAE_MCMC_inversion_example_fwi.png}b). 

The inverted mean model and standard deviation model are presented in Figure~\ref{fig:fine_tune/2025-08-05-09-18-39CAE_MCMC_inversion_example_fwi.png}g and \ref{fig:fine_tune/2025-08-05-09-18-39CAE_MCMC_inversion_example_fwi.png}h respectively. Although the pre-trained CAE-based Bayesian inversion is still able to recover the low-wavenumber background of the model, the inverted mean model exhibits pronounced network-induced artifacts. This degradation arises because the pre-trained CAE fails to adequately represent the geological features present in the Marmousi model. These results highlight the necessity of incorporating a fine-tuning step to adapt the network for out-of-distribution inversion scenarios.

\subsection{Out-of-distribution inversion through fine-tuning}

As discussed previously, when the subsurface model to be inverted exhibits features that deviate significantly from those present in the training datasets, the inversion might struggle to recover a model that adequately explains the observed data. To address this challenge, we propose a transfer learning scheme based on online fine-tuning.

\begin{figure*}[!t]
    \centering
    \includegraphics[width=0.7\linewidth]{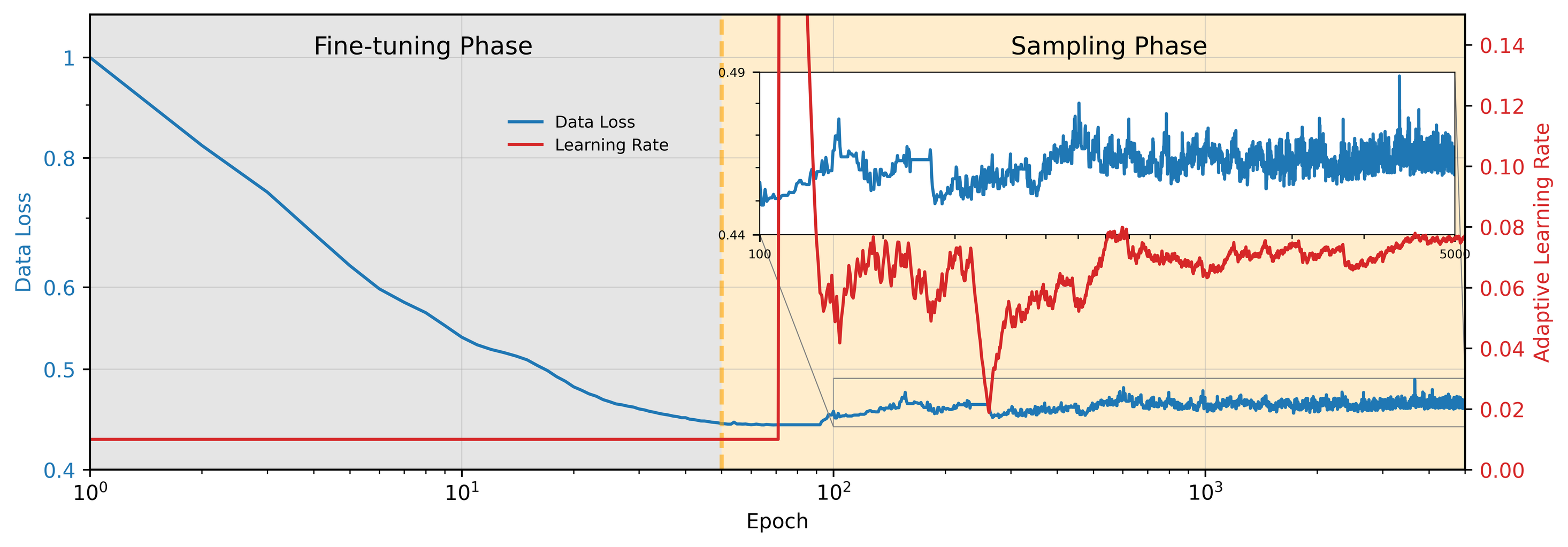}
    \caption{We show the data loss (blue curve) and adaptive learning rate (red curve) during the inversion. The inversion starts from the fine-tuning phase (gray shaded), where only the decoder parameters are updated. In the sampling phase (yellow shaded), the decoder parameters are frozen and the learning rate is automatically updated using adaptive MCMC algorithm.}
    \label{fig:fine_tune/data_loss_plot_with_adaptive_lr.png}
\end{figure*}

In contrast to the in-distribution case, where the inversion proceeds directly to the sampling phase, the out-of-distribution inversion begins with a fine-tuning phase. Figure~\ref{fig:fine_tune/data_loss_plot_with_adaptive_lr.png} illustrates the evolution of data misfit during the inversion. The fine-tuning phase, indicated by the gray-shaded region, runs for 50 iterations and demonstrates a consistent reduction in data loss. This is followed by the sampling phase (yellow-shaded), during which we employ an adaptive gradient-based MCMC algorithm to efficiently draw posterior samples. The learning rate, shown as a red curve in the same figure, is automatically adjusted throughout the inversion to achieve optimal sampling performance.

\begin{figure*}[!t]
    \centering
    \includegraphics[width=0.8\linewidth]{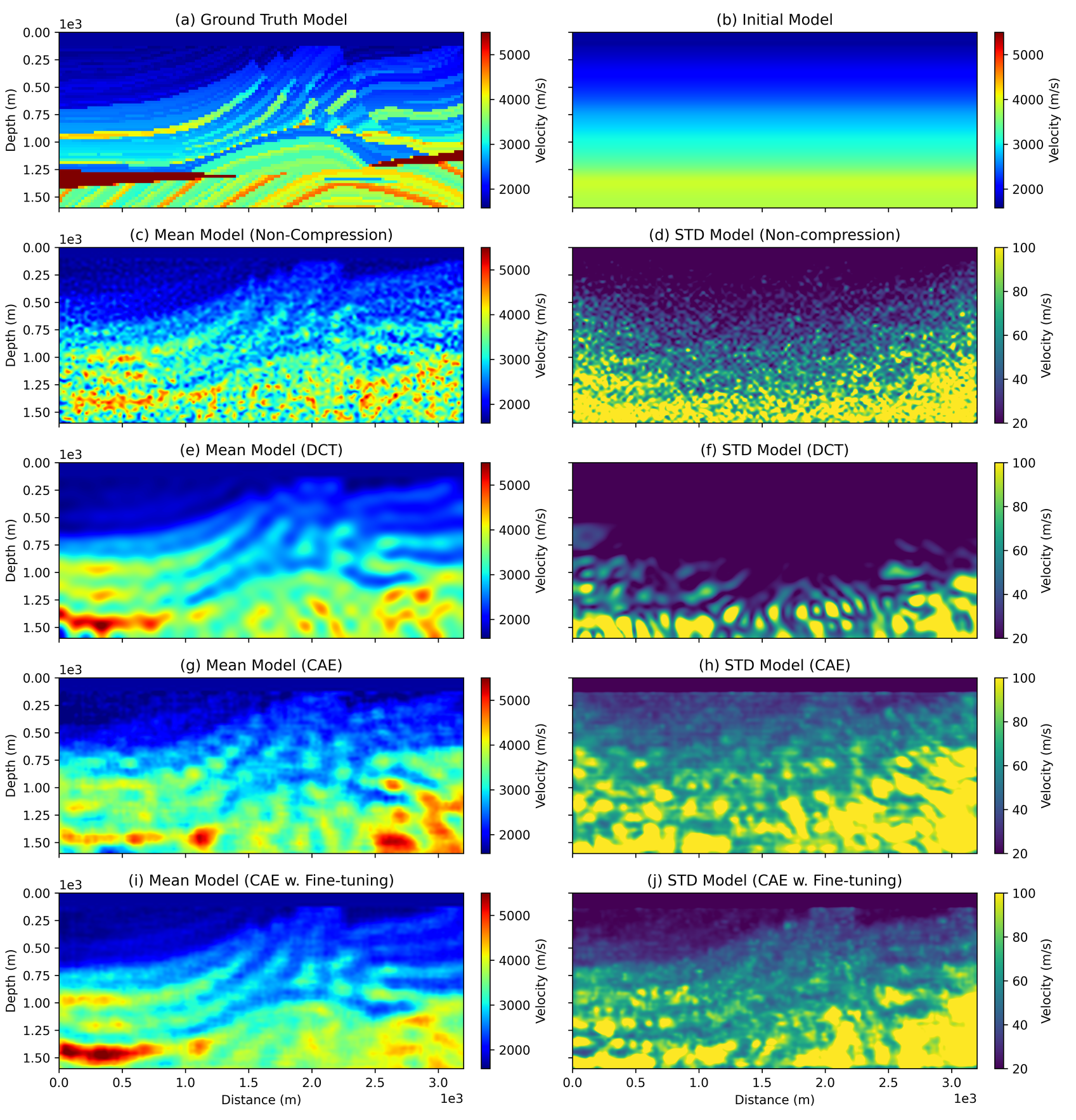}
    \caption{We applied the proposed method to the Marmousi model (a), which has very different characteristics with the training dataset. The inversion starts from a 1D gradient velocity model (b). We compare different methods including non-compression method (c \& d), DCT-based method (e \& f), CAE-based method without fine-tuning (g \& h) and CAE-based method with fine-tuning (i \& j). We show inverted mean models (c, e, g \& i) and standard deviation models (d, f, h, \& j) from different methods. All methods were using the same adaptive MCMC sampling algorithm, starting from the same 1D initial model, using the same synthetic observed data, and all ran for same 5,000 iterations.}
    \label{fig:fine_tune/2025-08-05-09-18-39CAE_MCMC_inversion_example_fwi.png}
\end{figure*}

In this example, we set the burn-in phase to \( N_{\text{burn-in}} = 500 \). Posterior samples beyond the burn-in period are used to compute the mean model (Figure~\ref{fig:fine_tune/2025-08-05-09-18-39CAE_MCMC_inversion_example_fwi.png}i) and the standard deviation model (Figure~\ref{fig:fine_tune/2025-08-05-09-18-39CAE_MCMC_inversion_example_fwi.png}j). The inverted mean model successfully recovers the low-wavenumber structure of the true model. The STD model reflects the estimated uncertainty of the inversion. As observed in the in-distribution case, higher uncertainty is present in poorly illuminated regions, particularly at the lower corners of the model. Additionally, uncertainty increases with depth, indicating reduced resolution in deeper regions due to limited acquisition coverage.

\begin{figure*}[!t]
    \centering
    \includegraphics[width=0.8\linewidth]{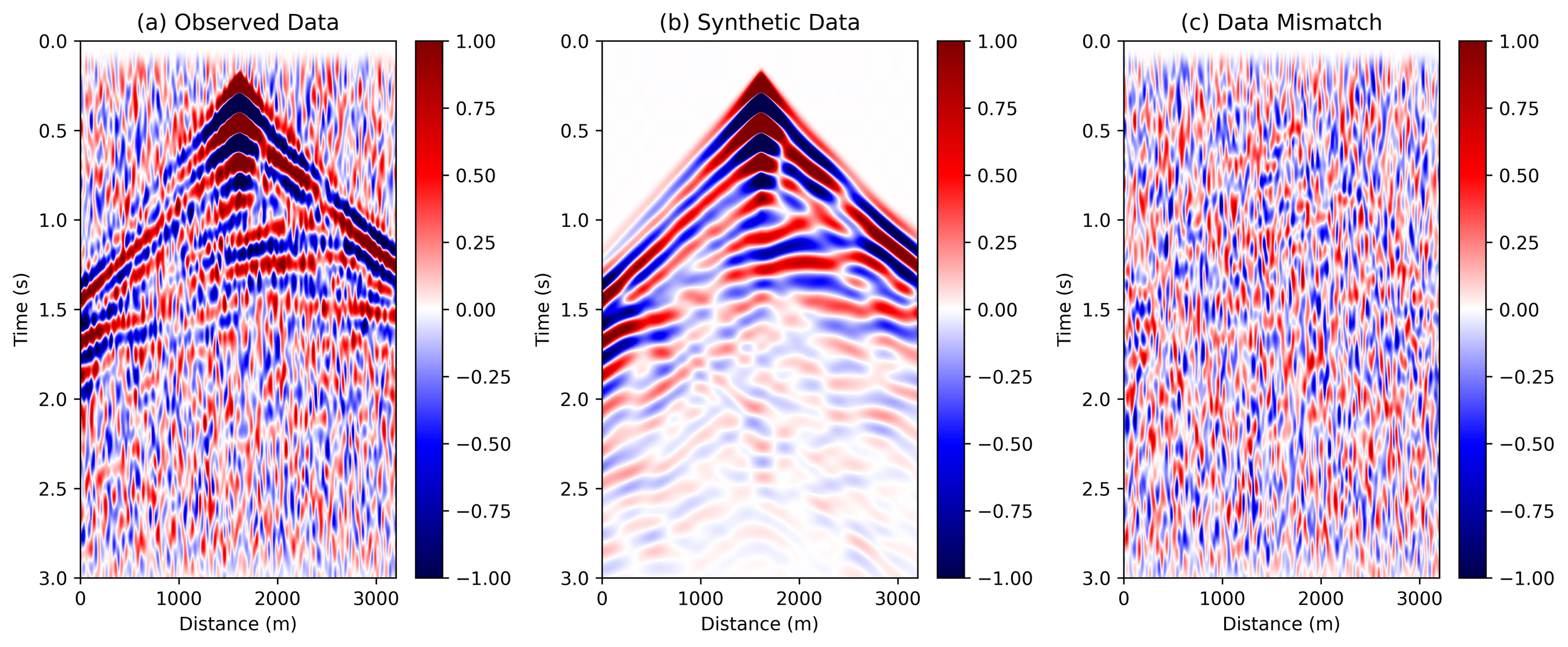}
    \caption{Comparison of noisy observed data (a), synthetic data using final inverted model (b), and the data mismatch (c).}
    \label{fig:fine_tune/2025-08-05-09-18-39_observed_simulated_data.png}
\end{figure*}

To benchmark the performance of the proposed method, we also present inversion results obtained using two alternative approaches: (i) adaptive gradient-based MCMC sampling in the full model space without dimensionality reduction (Figure~\ref{fig:fine_tune/2025-08-05-09-18-39_observed_simulated_data.png}c--d), and (ii) DCT-based dimensionality reduction (Figure~\ref{fig:fine_tune/2025-08-05-09-18-39_observed_simulated_data.png}e--f). For a fair comparison, all methods are applied to the same synthetic dataset, employ the same adaptive MCMC sampling algorithm, start from the same 1D initial velocity model, and run for an equal number of 5,000 iterations.

Without dimensionality reduction, the conventional fine-grid sampling-based Bayesian inversion fails to recover a meaningful velocity model. The inverted mean model is dominated by high-frequency noise, primarily caused by the stochastic perturbations introduced in the high-dimensional grid-based parameterization. The corresponding standard deviation model is also heavily contaminated by noise, indicating poor chain convergence. This behavior highlights the well-known difficulty of sampling in high-dimensional spaces, where an impractically large number of samples is required for reliable inference. In practice, more than $10^5$ iterations would likely be needed to achieve inference for such problems.

In contrast, the DCT-based method significantly improves the inversion results by reducing the effective dimensionality of the model space. The reconstructed velocity model captures the overall large-scale structure reasonably well. However, a characteristic DCT footprint is still visible in the inverted mean model, particularly in regions with higher uncertainty. Moreover, the corresponding STD model exhibits biased uncertainty estimates, with unrealistically low variance in the mid-to-upper regions of the model. Pronounced stripe-like artifacts are also observed, which can be attributed to the global nature of the DCT basis functions and their limited ability to represent localized geological features. Overall, this comparison highlights the importance of dimensionality reduction in Bayesian FWI and demonstrates the advantages of data-driven approaches. Compared to analytical transforms such as DCT, learned representations such as CAEs provide more flexible and geologically consistent parameterizations, leading to improved inversion quality and more reliable uncertainty quantification.



\begin{figure*}[!t]
    \centering
    \includegraphics[width=0.9\linewidth]{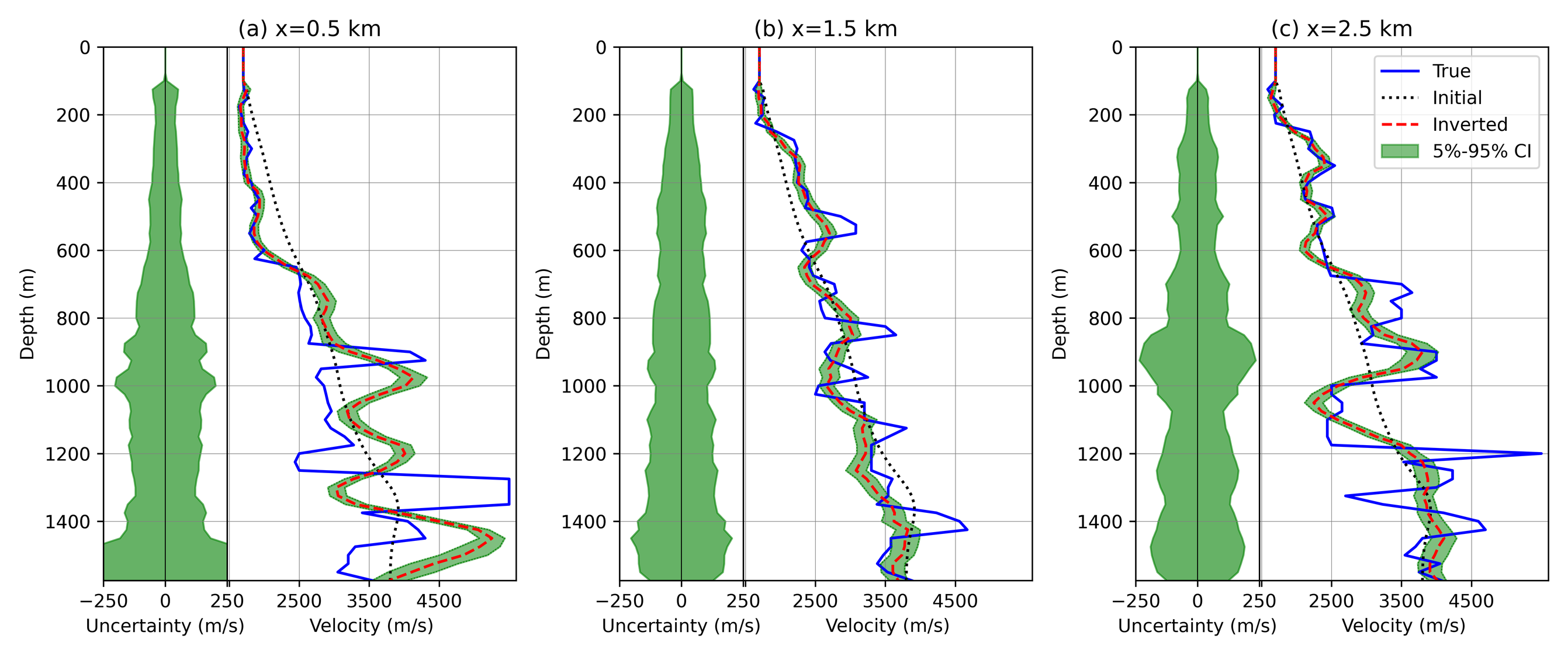}
    \caption{We show the trace plot at three different horizontal locations (a) 0.5 km, (b) 1.5 km and (c) 2.5 km, comparing the true model (blue), initial model (gray dotted), inverted model (red dashed), as well as the confidence interval between 5th and 95th percentiles (green shaded).}
    \label{fig:fine_tune/fine_tune_CAE_MCMC_inversion_traceplot.png}
\end{figure*}

For CAE-based Bayesian inversion, we also compare the synthetic data generated using the inverted mean model (Figure~\ref{fig:fine_tune/2025-08-05-09-18-39_observed_simulated_data.png}b) with the observed data. The data residual, shown in Figure~\ref{fig:fine_tune/2025-08-05-09-18-39_observed_simulated_data.png}c, primarily consists of noise, indicating a good match between the simulated and observed wavefields. This validates the accuracy of the inverted mean model. Figure~\ref{fig:fine_tune/fine_tune_CAE_MCMC_inversion_traceplot.png} presents trace plots and CI plot at three different horizontal locations. The inverted mean model (red dashed line) provides a good low-wavenumber approximation of the true model (blue line). The shaded confidence interval (green shaded), bounded by the 5th and 95th percentiles, quantifies the uncertainty resulting from the frequency-limited inversion.


\subsection{Extension to higher frequency}

\begin{figure*}[!t]
    \centering
    \includegraphics[width=0.8\linewidth]{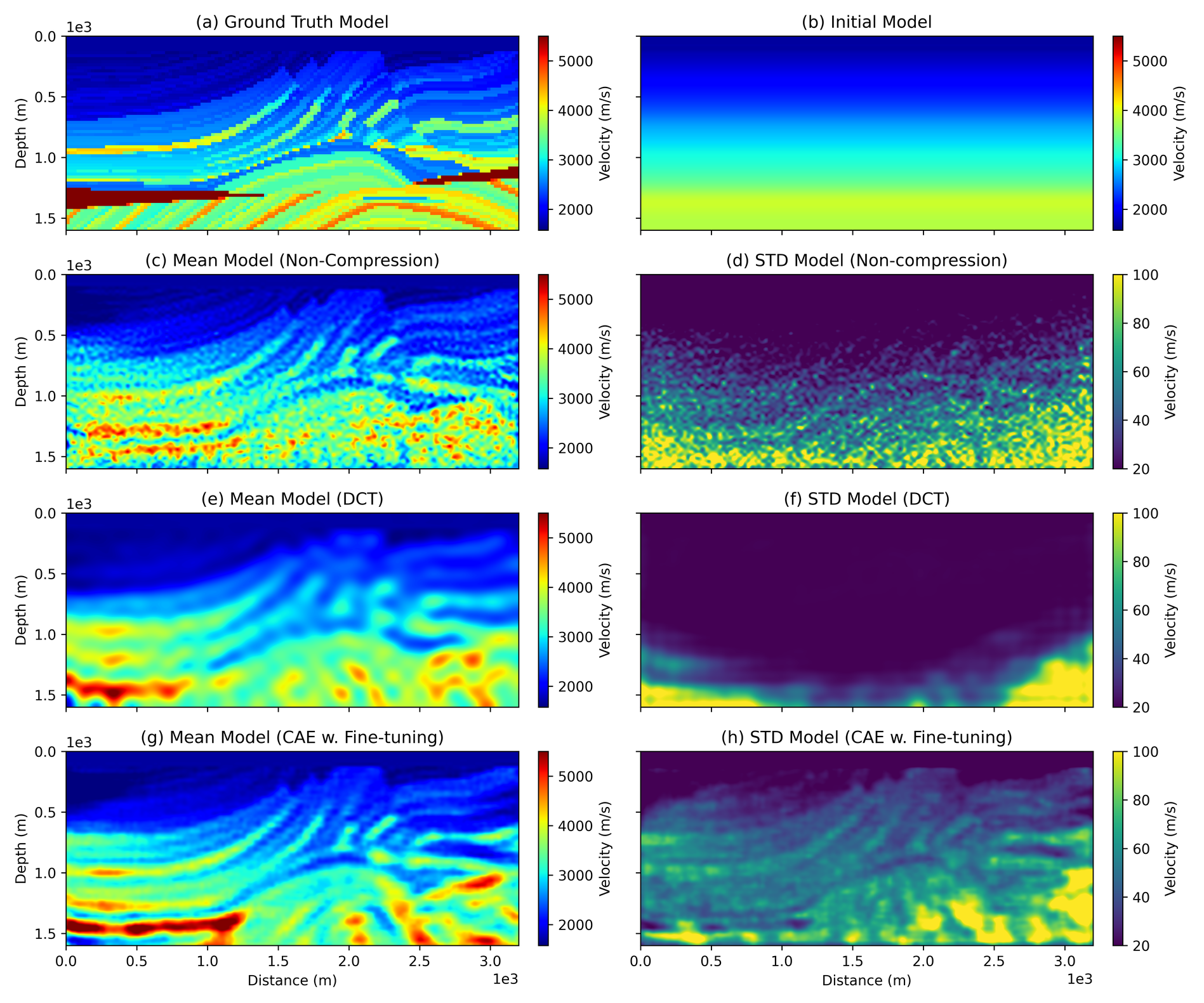}
    \caption{The proposed method was verified with a multi-scale strategy with data frequency up to 15 Hz in second stage. The inverted models from the first-stage 5Hz inversion (Figure \ref{fig:fine_tune/2025-08-05-09-18-39CAE_MCMC_inversion_example_fwi.png}) were used as prior model for each method. The ground truth model was shown in (a) and the inversion starts from the 1D gradient velocity model (b). Similar to the first-stage inversion, different methods are compared, including non-compression method (c \& d), DCT-based method (e \& f) and CAE-based method with fine-tuning (g \& h). We show inverted mean models (c, e \& g) and standard deviation models (d, f \& h) from different methods. All methods were using the same adaptive MCMC sampling algorithm, starting from the same 1D initial model, using the same synthetic observed data filtered with cut-off frequency of 15 Hz, and all ran for same 5,000 iterations. We also apply the same model covariance matrix to all methods, i.e. $\mathbf{C_m}=\beta\mathbf{I}$, with $\beta$ optimum tuned.}    \label{fig:rev2/MCMC_inversion_example_fwi_all4comparision_paper.png}
\end{figure*}

\begin{figure*}[!t]
    \centering
    \includegraphics[width=0.8\linewidth]{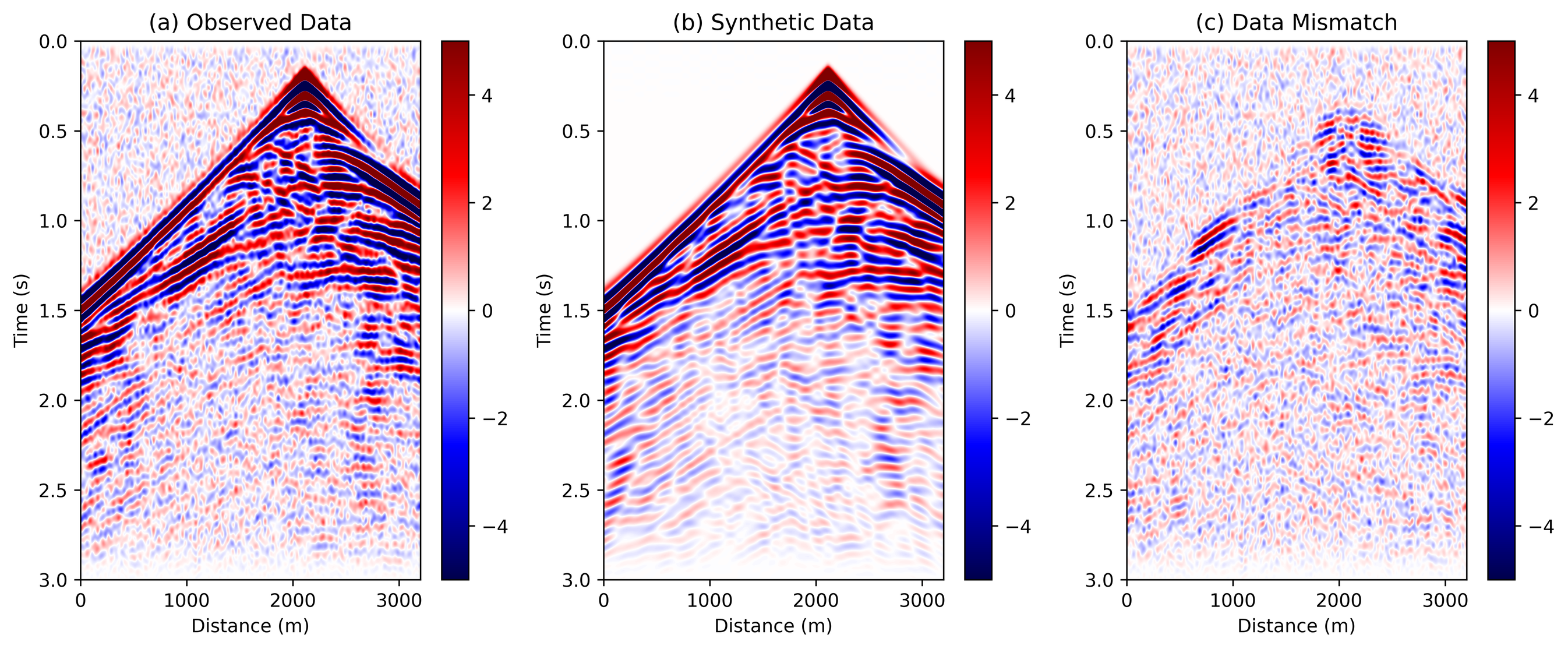}
    \caption{Comparison of noisy observed data (a), synthetic data using final inverted model using CAE with fine-tuning (b), both filtered to 15 Hz, and the data mismatch (c).}
    \label{fig:Fig/rev2/2026-06-22-00-28-10_observed_simulated_data.png}
\end{figure*}

\begin{figure*}[!t]
    \centering
    \includegraphics[width=0.7\linewidth]{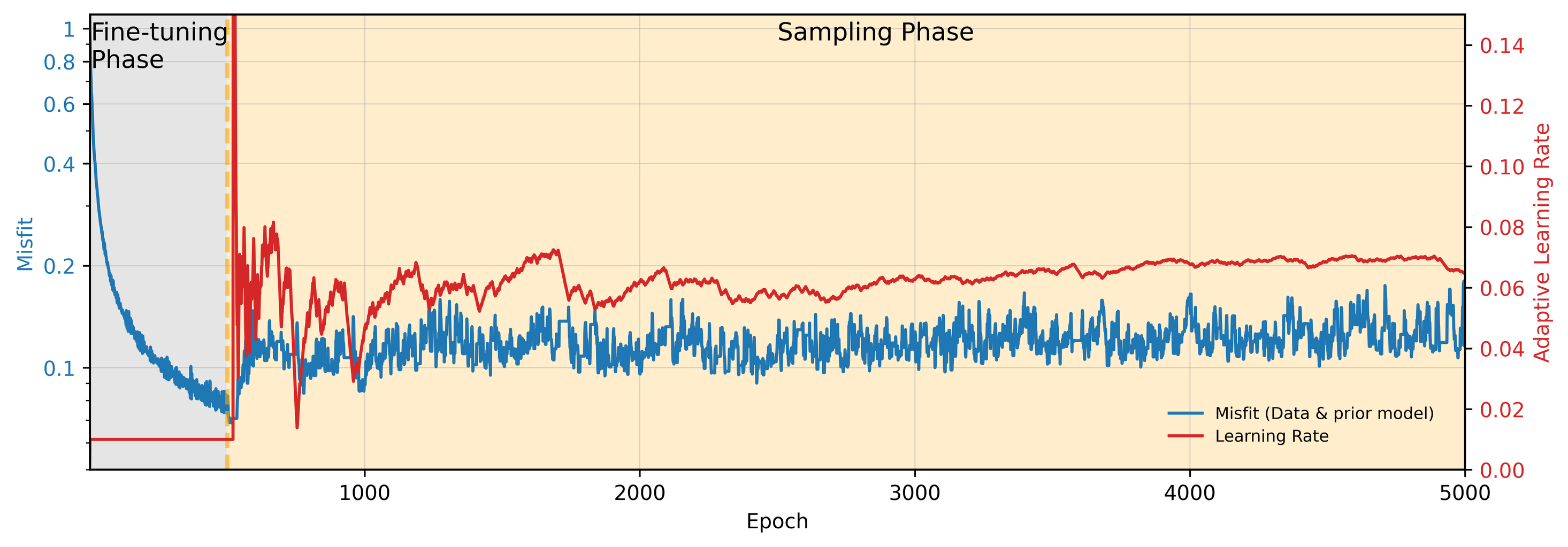}
    \caption{We show the overall misfit (blue curve) and adaptive learning rate (red curve) during the second-stage inversion for the Marmousi model example. Similarly to the first stage, the inversion starts from the fine-tuning phase (gray shaded), where only the decoder parameters are updated. In the sampling phase (yellow shaded), the decoder parameters are frozen and the learning rate is automatically updated using adaptive MCMC algorithm.}    \label{fig:Fig/rev2/data_loss_plot_with_adaptive_lr.png}
\end{figure*}

\begin{figure*}[!t]
    \centering
    \includegraphics[width=0.9\linewidth]{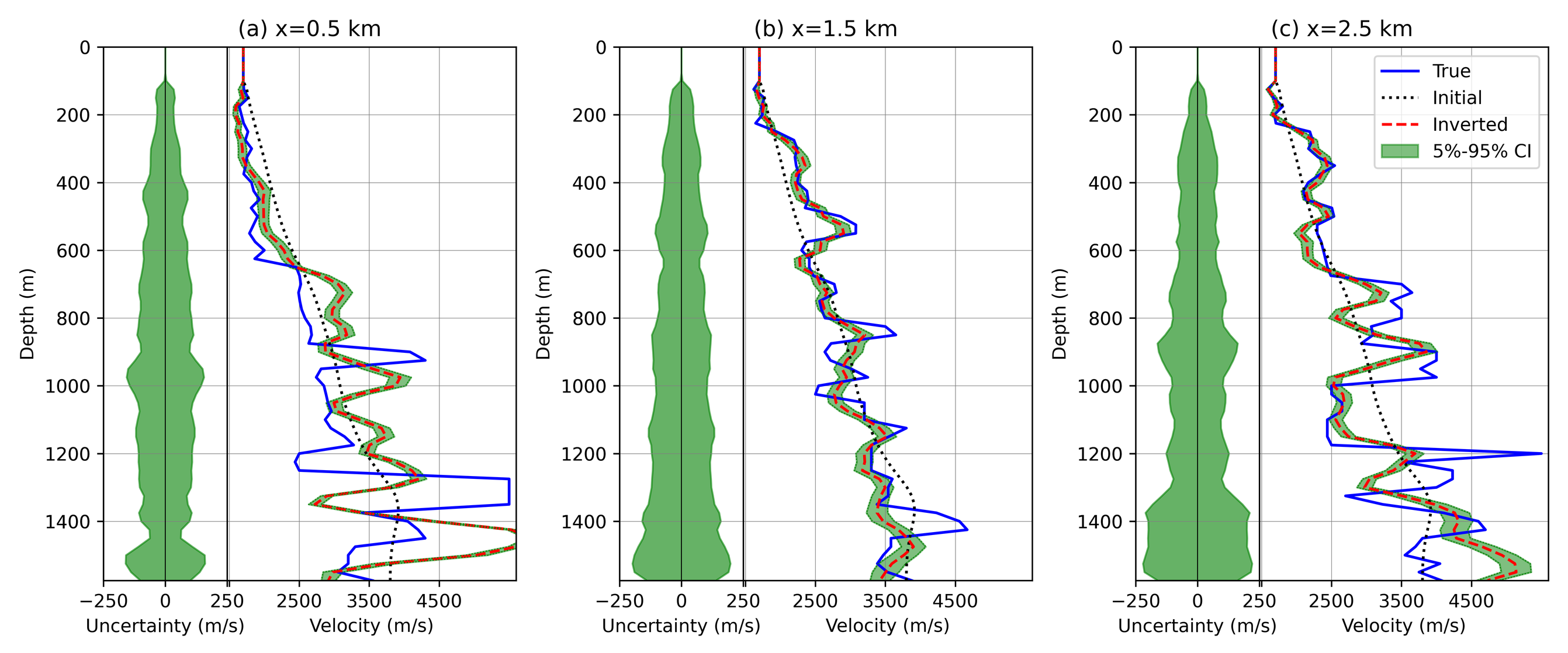}
    \caption{We show trace plots at three different horizontal locations (a) 0.5 km, (b) 1.5 km and (c) 2.5 km extracted from Figure (\ref{fig:rev2/MCMC_inversion_example_fwi_all4comparision_paper.png}). We compare the true model (blue), initial model (gray dotted), inverted model (red dashed), as well as the confidence interval between 5th and 95th percentiles (green shaded).}    \label{fig:rev2/Marmousi_fine_tune_CAE_MCMC_inversion_traceplot_ribbon.png}
\end{figure*}

The numerical examples presented in the previous section demonstrate the effectiveness of the proposed framework using seismic data with frequency components up to 5\,Hz for out-of-distribution inversion. Although low-frequency data are essential for recovering the large-scale, low-wavenumber structures of the subsurface and mitigating cycle-skipping, they inherently limit the achievable resolution of the inversion. In practical FWI workflows, a multi-scale inversion strategy is often adopted, where progressively higher-frequency information is introduced after a reliable background model has been established. Following this philosophy, we further evaluate the proposed method for out-of-distribution inversion using a two-stage inversion strategy. In the first stage, inversion is performed using data filtered to 5\,Hz, and the corresponding inversion results are presented in the previous section. In the second stage, the frequency content is extended to 15\,Hz, allowing the inversion to recover finer-scale geological features. For each method, the inverted mean model obtained from the first-stage 5\,Hz inversion is used as the prior model for the second-stage inversion. The prior covariance matrix is defined as
\begin{equation}
    \mathbf{C}_{m} = \beta \mathbf{I},
\end{equation}
where $\mathbf{I}$ denotes the identity matrix and $\beta$ is a tuning parameter controlling the influence of the prior model. All methods employ the same adaptive gradient-based MCMC algorithm, use the same observed dataset filtered to 15\,Hz, and are run for 5,000 iterations to ensure a fair comparison.

Figure~\ref{fig:rev2/MCMC_inversion_example_fwi_all4comparision_paper.png} shows the inversion results obtained from the second-stage inversion. The non-compression approach (Figure~\ref{fig:rev2/MCMC_inversion_example_fwi_all4comparision_paper.png}c--d) recovers additional high-wavenumber information compared to the corresponding 5\,Hz inversion (Figure~\ref{fig:fine_tune/2025-08-05-09-18-39CAE_MCMC_inversion_example_fwi.png}c--d). However, the reconstructed model is heavily contaminated by grid-scale Gaussian noise, and the estimated standard deviation model exhibits significant fluctuations. This behavior is expected for a high-dimensional Bayesian inversion problem, where a substantially larger number of posterior samples is required to adequately explore the posterior distribution and achieve stable uncertainty estimates. The DCT-based dimensionality reduction method (Figure~\ref{fig:rev2/MCMC_inversion_example_fwi_all4comparision_paper.png}e--f) produces a smoother inversion result with improved convergence. Nevertheless, the introduction of higher-frequency data does not lead to a substantial increase in resolution compared to the 5\,Hz DCT inversion (Figure~\ref{fig:fine_tune/2025-08-05-09-18-39CAE_MCMC_inversion_example_fwi.png}e--f). This observation is primarily attributed to the limited number of retained DCT coefficients and the inherent inability of the truncated DCT basis to effectively represent complex high-wavenumber geological structures. As a result, the inversion remains constrained to relatively smooth models despite the availability of higher-frequency information.

Among the three approaches, the CAE-based method with online fine-tuning (Figure~\ref{fig:rev2/MCMC_inversion_example_fwi_all4comparision_paper.png}g--h) delivers the highest-resolution inversion result. The recovered mean model successfully reconstructs many of the fine-scale structural features present in the Marmousi model while maintaining geological consistency. Furthermore, the corresponding standard deviation model reveals a spatially meaningful uncertainty distribution that is correlated with the acquisition geometry and model complexity. Compared with the standard deviation models obtained from the direct CAE inversion at 5\,Hz (Figure~\ref{fig:fine_tune/2025-08-05-09-18-39CAE_MCMC_inversion_example_fwi.png}h) and the fine-tuned CAE inversion at 5\,Hz (Figure~\ref{fig:fine_tune/2025-08-05-09-18-39CAE_MCMC_inversion_example_fwi.png}j), the uncertainty level is further reduced after introducing the 15\,Hz data. This trend indicates that both the online fine-tuning strategy and the incorporation of higher-frequency information contribute to a more informative likelihood function and improved posterior constraint. Consequently, the inversion results become increasingly confident as additional information is incorporated, leading to enhanced model resolution and reduced posterior uncertainty.

To further evaluate the quality of the inversion, we compare the observed data with the synthetic data generated from the final inverted mean model in Figure~\ref{fig:Fig/rev2/2026-06-22-00-28-10_observed_simulated_data.png}. Both datasets are filtered to 15\,Hz. As shown in Figure~\ref{fig:Fig/rev2/2026-06-22-00-28-10_observed_simulated_data.png}a--b, the synthetic data successfully reproduce the major kinematic and amplitude characteristics of the observed data. The corresponding data residual (Figure~\ref{fig:Fig/rev2/2026-06-22-00-28-10_observed_simulated_data.png}c) is largely dominated by random noise, with only limited coherent events remaining. This indicates that the recovered model explains most of the information contained in the observed data and confirms the effectiveness of the proposed inversion framework at higher frequencies. Figure~\ref{fig:Fig/rev2/data_loss_plot_with_adaptive_lr.png} shows the evolution of the overall objective function and adaptive learning rate during the second-stage inversion. Similar to the first-stage inversion, the inversion begins with a short fine-tuning phase (500 epochs), during which only the decoder parameters are updated. The data misfit decreases rapidly during this stage, indicating that the decoder is successfully adapted to the Marmousi-style geological features. Subsequently, the inversion enters the sampling phase, where the decoder parameters are frozen and posterior samples are generated using the adaptive gradient-based MCMC algorithm. The objective function remains stable throughout the sampling phase, while the adaptive learning rate automatically adjusts to maintain efficient posterior exploration.

Figure~\ref{fig:rev2/Marmousi_fine_tune_CAE_MCMC_inversion_traceplot_ribbon.png} presents vertical trace comparisons at three horizontal locations extracted from the inversion results. The inverted mean model closely follows the true velocity profile and significantly improves upon the initial model. Compared with the corresponding 5\,Hz inversion results, the uncertainty intervals are noticeably reduced. This observation is consistent with the reduction of the standard deviation observed in Figure~\ref{fig:rev2/MCMC_inversion_example_fwi_all4comparision_paper.png}h and further demonstrates that the incorporation of higher-frequency information leads to a more constrained posterior distribution.

Overall, the multi-scale Marmousi example demonstrates the advantages of the proposed CAE-based dimensionality reduction framework over both the non-compression and DCT-based approaches. While the non-compression method suffers from the curse of dimensionality and requires substantially more samples to achieve convergence, the DCT-based method is fundamentally limited by the expressiveness of its predefined basis functions. As a result, high-frequency information cannot be effectively translated into corresponding high-wavenumber model updates. In contrast, the CAE learns a data-driven latent representation, enabling a more efficient parameterization of the subsurface and a more effective utilization of the information contained in the seismic data. Combined with the online fine-tuning strategy, the learned latent space can be adapted to out-of-distribution settings, allowing the inversion to recover both large-scale and fine-scale structural features. The resulting posterior models exhibit higher resolution and reliable uncertainty estimates, highlighting the potential of learned priors for large-scale Bayesian full waveform inversion.

\section{DISCUSSION}

We demonstrates the potential of convolutional autoencoders for efficient dimensionality reduction in sampling based Bayesian inference for seismic full waveform inversion. We show that the proposed method can efficiently estimate posterior distribution. Compared to traditional methods such as the discrete cosine transform, our results show that even a relatively simple CAE architecture consistently outperforms DCT in preserving low-wavenumber features when constrained to the same latent dimensionality. This advantage is particularly beneficial for downstream tasks such as Bayesian inversion, where high-fidelity reconstructions of subsurface models and efficient sampling from the posterior distribution are essential.

\subsection{Training dataset}

However, the effectiveness of CAE-based approaches critically depends on the availability and quality of training data. The learned CAE prior inherently reflects the statistical characteristics of the training dataset, which may introduce bias when applied to geological settings not represented in the data. Unlike analytical transforms such as DCT, CAE requires a representative training dataset that captures the variability and complexity of real-world subsurface structures. Constructing such datasets remains a challenge, especially in practical scenarios where ground-truth velocity models are limited. In Bayesian FWI, such prior bias can overly constrain the posterior distribution, potentially steering the inversion toward structures favored by the training dataset and limiting the ability to recover out-of-distribution subsurface features.
This limitation emphasizes the need for collaborative efforts in building large-scale, diverse velocity model datasets that can better serve as reliable training datasets.

An emerging trend that may address this limitation is the use of generative models, such as generative adversarial networks (GANs, \cite{goodfellow2014generative}) or diffusion models \cite{ho2020denoising}, to synthesize high-quality and geologically plausible velocity models. For example, \cite{puzyrev2022geophysical} propose to use GAN as a generator for 2D subsurface models based on realistic density and stratigraphy models. \cite{wang2024controllable} employ conditional generative diffusion models for seismic velocity generation, in which they show diffusion models provide a flexible framework to incorporate the prior knowledge, such as well logs and reflectivity images. These models can be trained on existing datasets and used to generate an effective set of diverse samples that respect geological realism. This capability can augment traditional handcrafted datasets, and help improve the robustness and generalization of CAE training.

Looking forward, we envision the development of subsurface foundation models for subsurface velocity structures, trained on a broad corpus of synthetic and real subsurface model data. Recent advances in seismic foundation model \cite{sheng2025seismic,dou2025geologicalmodel3dpromptable} highlight the capability of foundation models in capturing all-purpose seismic features and applicability to various tasks. We expect a subsurface foundation model could provide a robust latent representation of subsurface features and significantly improve the generalization capacity of CAE-based methods across diverse geological contexts.

\subsection{Resolution improvement}

Another pitfall lies in the fact that CAE inherently suffers from over-smoothed reconstructions due to its reliance on convolutional filters and bottleneck structures. This limitation is evident in the numerical examples presented in this study, where the reconstructed velocity models capture the large-scale, low-wavenumber structures well but lack high-resolution details. 
Such smoothing effects may limit the ability of the CAE to represent sharp geological features, which are important for accurate subsurface characterization.

To overcome this limitation, more advanced generative models have been proposed in the machine learning literature. For instance, hybrid architectures such as VAE-GAN \cite{larsen2016autoencoding} combine the structured latent space of variational autoencoders (VAE) with the adversarial training of generative adversarial networks (GAN), enabling both compact representation and high-fidelity reconstruction. 
More recently, latent diffusion models \cite{rombach2022high} have demonstrated remarkable capability in generating high-resolution images by performing diffusion processes in a compressed latent space, effectively balancing efficiency and reconstruction quality. 
These models offer a promising direction for improving the representational power of learned priors in geophysical inversion, as they can potentially preserve fine-scale geological features while maintaining a tractable latent space for sampling-based Bayesian inference.

\subsection{Transfer learning}

One potential avenue to further improve generalization is the use of transfer learning. By fine-tuning a pre-trained CAE on a set of task-specific or region-specific velocity models, we can adapt the model to new geological settings without retraining. Our results on out-of-distribution (OOD) inversion validate this approach, where online fine-tuning enables accurate reconstructions even when the test model deviates significantly from the training distribution. To mitigate overfitting to the new dataset, we restrict the number of fine-tuning iterations. In addition, although not explored in this work, alternative strategies commonly used in machine learning—such as freezing a subset of network parameters—could be employed to further improve generalization during fine-tuning.

\subsection{MCMC configuration}

Although only single MCMC chain is employed in the numerical examples presented in this work, the proposed framework is not restricted to either single-chain or multi-chain configurations. Using multiple chains offers several advantages, including multi-chain convergence diagnostics, improved exploration of the posterior distribution through diverse initializations, more efficient utilization of computational resources, and the potential incorporation of advanced strategies such as parallel tempering (e.g.\cite{hu2024dct}). In principle, initializing chains from different regions of the model space can enhance robustness and reduce the risk of insufficient exploration. However, for high-dimensional FWI problems, it is often beneficial to initialize sampling from a reasonable background velocity model to ensure stable convergence and physically meaningful updates.

\subsection{Field data application}

A practical limitation of the proposed CAE-based framework is that the network is trained on velocity models with fixed spatial dimensions and grid resolution. This constraint is common in convolutional neural network (CNN)-based inversion methods, where the architecture is typically designed for a predefined input size. 
As a result, direct application to inversion problems with different domain sizes or grid discretizations is not straightforward. 
In general, dimensional adaptation can only be performed outside the network, either before the encoder or after the decoder. 

To mitigate this limitation, one possible strategy is to train the CAE on datasets with sufficiently large spatial coverage. This enables the trained model to be applied to smaller domains through cropping, or to different grid sizes through appropriate resizing. In such cases, the reconstructed models can be interpolated or resampled to match the target inversion grid, provided that the physical scaling is handled carefully to avoid distortion of geological features. Nevertheless, when there is a significant mismatch between the training and target domains, retraining of the network may still be required. Addressing this limitation through more flexible architectures or scale-adaptive training strategies remains an important direction for future work.

In summary, CAEs offer a powerful framework for data-driven model compression in geophysical applications, future work might focus on improving training data diversity, exploring transfer learning strategies, and building more generalizable foundation models. These directions hold promise for making CAE-based inversion methods applicable to real-world scenarios.

\section{CONCLUSIONS}

In this paper, we present a deep convolutional autoencoder framework as a learned prior for Bayesian FWI. The CAE architecture, consisting of two convolutional layers in both the encoder and decoder, enables efficient dimensionality reduction for velocity models while preserving essential structural features. Leveraging the pre-trained CAE, we introduced an adaptive gradient-based MCMC sampling method in the latent space to efficiently explore the posterior distribution. Compared to traditional transform-based representations, the proposed method better captures complex geological structures and provides a more expressive, data-driven prior. Synthetic experiments demonstrate that our approach can efficiently estimate the posterior distribution, which can be used to quantify the uncertainties associated with the highly nonlinear FWI.

\appendices
\section{Experiment on latent dimension}
\label{sec:appendix_latent_dim}

To evaluate the impact of latent dimensionality on autoencoder training, we define a value ladder for the latent dimension as \( \bar{n}_j = 2^j \), for \( j = 3, \ldots, 7 \). Table~\ref{tab:training_time} summarizes the total number of trainable parameters (weights and biases) in the autoencoder, along with the corresponding training time for each configuration. All training experiments were conducted on a single NVIDIA RTX A5000 GPU.



\begin{figure}
    \centering
    \includegraphics[width=\linewidth]{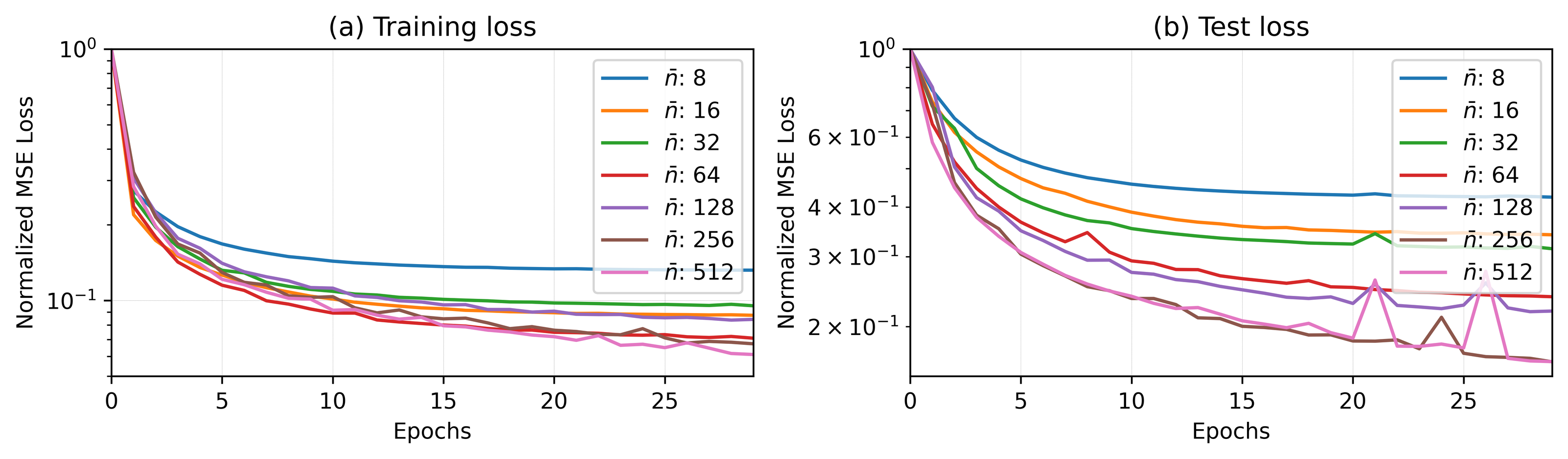}
    \caption{Training and test loss for the designed CAE with different latent dimensions.}
    \label{fig:training_loss}
\end{figure}

The training process runs for 30 epochs. Figure~\ref{fig:training_loss} presents the training and validation loss curves for autoencoders trained with different latent dimensions. As the latent dimension increases, the final training loss decreases, indicating that larger latent spaces are more effective at capturing the underlying features of the dataset. A similar trend is observed for the validation loss. However, when the latent dimension is small (e.g., \( \bar{n} < 64 \)), the validation loss plateaus around 0.4. In contrast, models with higher latent dimensions (e.g., \( \bar{n} > 128 \)) achieve validation losses below 0.2. This suggests that low-dimensional autoencoders have limited generalization capability compared to those with higher latent capacity.

\begin{figure}
    \centering
    \includegraphics[width=\linewidth]{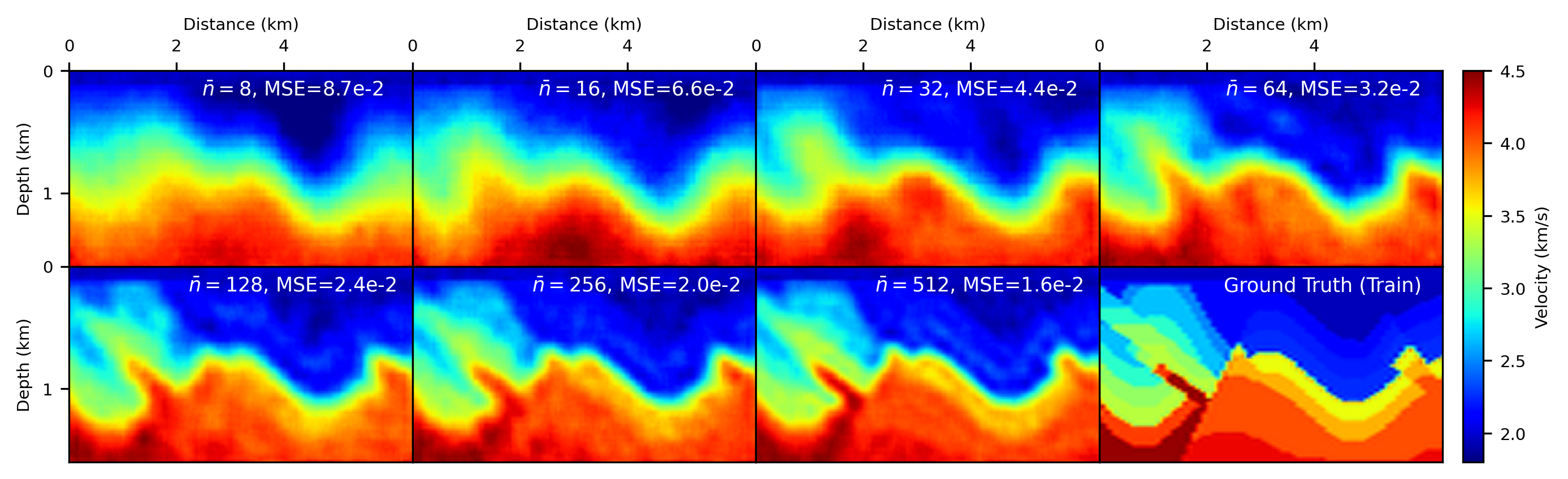}
    \caption{One sample from training dataset is selected to show the comparison of CAE compression for different latent dimensions.}
    \label{fig:training_dataset_verification}
\end{figure}

\begin{figure}
    \centering
    \includegraphics[width=\linewidth]{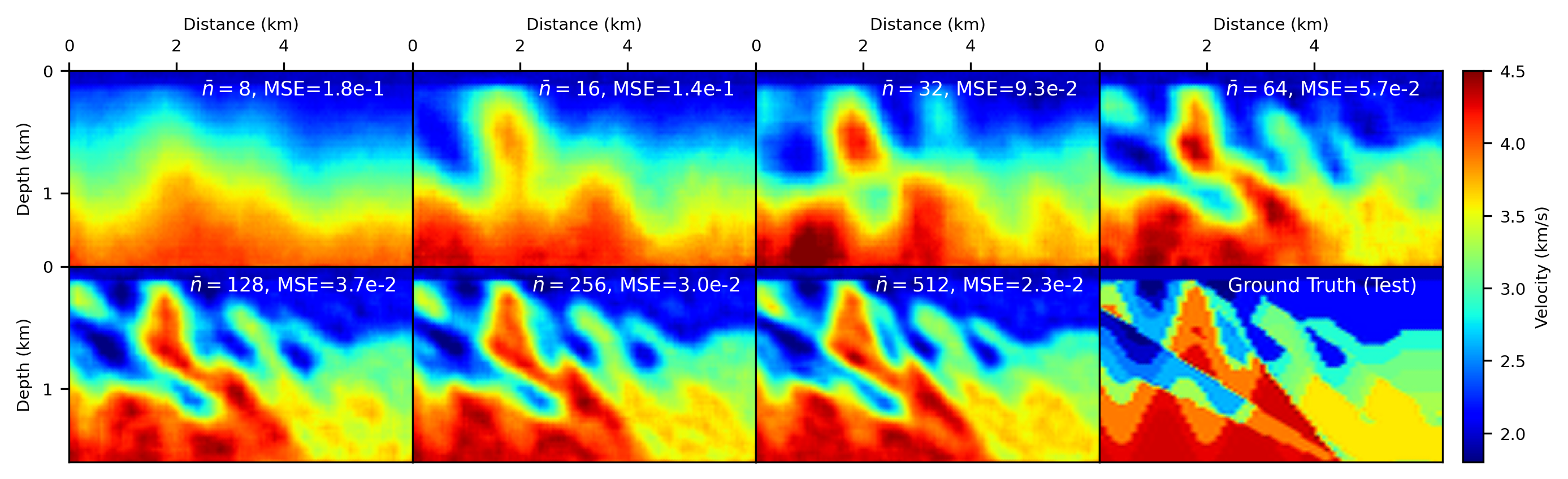}
    \caption{Same as in Figure \ref{fig:training_dataset_verification} but for a sample from test dataset.}
    \label{fig:test_dataset_verification}
\end{figure}

After training, we evaluate the reconstruction performance of the autoencoder on two representative velocity models: one from the training set (Figure~\ref{fig:training_dataset_verification}) and one from the test set (Figure~\ref{fig:test_dataset_verification}). We observe that, for sufficiently large latent dimensions (e.g., \( \bar{n} > 128 \)), the autoencoder successfully captures the low-wavenumber structure of the velocity models, demonstrating its effectiveness in learning a compact and meaningful latent representation.

\begin{table}[]
    \centering
    \caption{Number of trainable parameters and training time of each autoencoder model with different latent dimensions.}
    \begin{tabular}{c|c|c}
        \hline
        Latent dimension $\bar{n}$ & No. of Trainable Parameters & Training time (s)  \\
        \hline
        8 & 594,633 & 194 \\
        16 & 1,118,929 & 196 \\
        32 & 2,167,521 & 197 \\
        64 & 4,264,705 & 199 \\
        128 & 8,459,073 & 202 \\
        256 & 16,847,809 & 209 \\
        512 & 33,625,281 & 223 \\
        \hline
    \end{tabular}
    \label{tab:training_time}
\end{table}

\section{Experiment on activation function}
\label{sec:appendix_activation}

Periodic activation functions have been shown to be effective in capturing fine-grained details of physical signals and in solving physics-informed problems (e.g., \cite{sitzmann2020implicit,song2021}). In this work, we incorporate the sine activation function into our autoencoder architecture and inversion framework to enhance their representational capacity. We compare the performance of the sine activation function with the conventional ReLU activation in both the autoencoder pre-training and the transfer learning inversion tasks. 

Table~\ref{tab:mse_activation_function} presents the mean squared error of the autoencoder predictions using different activation functions on both training and validation samples. The results indicate that for small latent dimensions, ReLU yields lower MSE than sine. However, as the latent dimension increases, sine activation consistently outperforms ReLU, resulting in lower reconstruction errors.

\begin{figure}
    \centering
    \includegraphics[width=\linewidth]{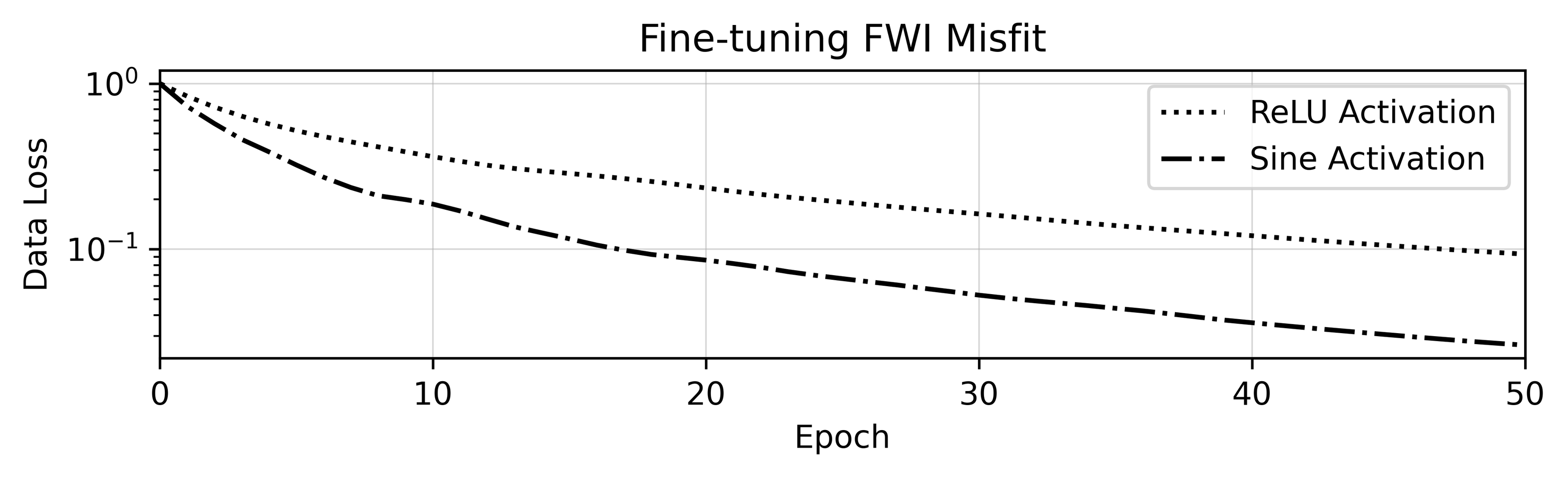}
    \caption{Comparison of fine-tuning phase data loss with different activation function in the autoencoder. }
    \label{fig:fine_tune_data_loss_log_comp_relu_sine}
\end{figure}

Further comparison of inversion performance is illustrated in Figure~\ref{fig:fine_tune_data_loss_log_comp_relu_sine}, which shows the evolution of data misfit during the fine-tuning phase. The sine activation function achieves significantly lower data loss compared to ReLU, confirming that periodic activations offer superior representation ability in the inversion process.

\begin{table}[]
    \centering
    \caption{MSE comparison of model predictions using the pre-trained autoencoder with different activation functions. The training case is the one shown in Figure \ref{fig:training_dataset_verification}, and the validation case is the one shown in Figure \ref{fig:2025-05-30__upsample_nearest_sine_DCT_comparison_vs_CAE_comparison_8_16_32_64_128.png}.}
    \begin{tabular}{c|c|c|c|c}
        \hline
         \multicolumn{1}{c}{Latent} & \multicolumn{2}{|c|}{Training MSE ($\times 10^{-2}$)} & \multicolumn{2}{|c}{Validation MSE ($\times 10^{-2}$)}  \\
         \cline{2-5}
        dimension $\bar{n}$ & ReLU & Sine  & ReLU & Sine \\
        \hline
        8 & \textbf{7.78}&	8.69&	14.45	&\textbf{14.38}\\
        16 &\textbf{6.03}&	6.57&	\textbf{12.54}	&12.91\\
        32 &4.45&	\textbf{4.37}&	\textbf{8.16}	&8.77\\
        64 &\textbf{2.94}&	3.16&	\textbf{5.50}	&5.69\\
        128 &2.50&	\textbf{2.38}&	4.67	&\textbf{4.22}\\
        256 &2.17&	\textbf{1.99}&	3.67	&\textbf{3.41}\\
        512 & 1.89&	\textbf{1.59}&	3.38 &	\textbf{2.68}\\
        \hline
    \end{tabular}
    \label{tab:mse_activation_function}
\end{table}

\section*{Acknowledgment}
The authors would like to express their sincere gratitude to the three anonymous reviewers and the Deputy Editor-in-Chief, X. Chen, for their careful review, valuable comments, and constructive suggestions, which have significantly improved the quality of this manuscript. Z. Zhao is supported by the Deep Earth Probe and Mineral Resources Exploration‐National Science and Technology Major Project of China (No. 2024ZD1001107), and the National Natural Science Foundation of China under Grants 42374138 and 2023002410.


\ifCLASSOPTIONcaptionsoff
  \newpage
\fi



\bibliographystyle{IEEEtran}
\bibliography{./bibtex/bib/IEEEabrv.bib,./bibtex/bib/IEEEexample.bib}
%







\begin{IEEEbiography}
[{\includegraphics[width=1in,height=1.25in,clip,keepaspectratio]{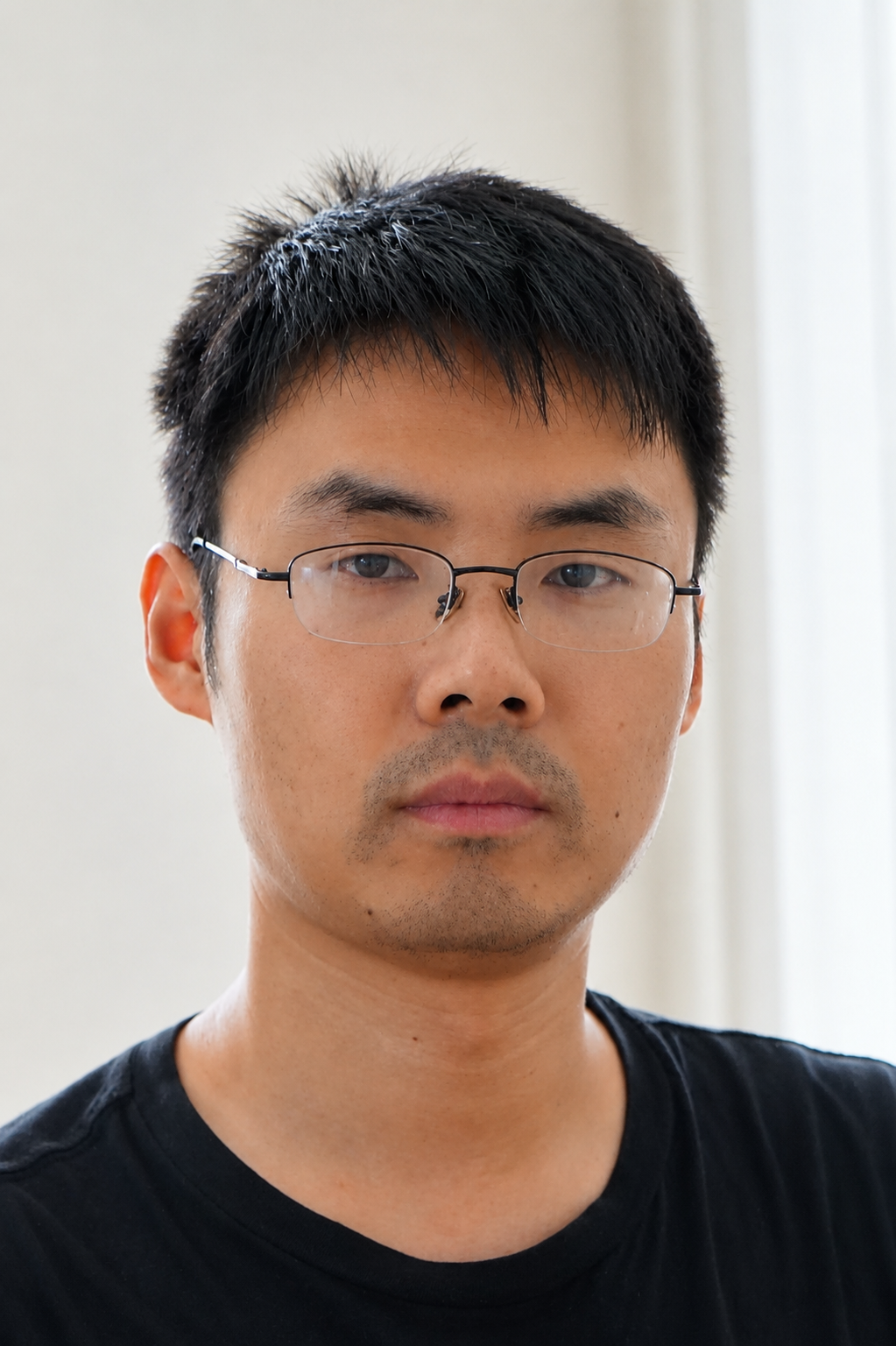}}]
{Shuhua Hu}
is currently a Senior Technologist/ Research Geophysicist at Halliburton Landmark. He earned a Bachelor’s degree in Exploration Technology \& Engineering from China University of Petroleum, Qingdao, China, and a Master’s degree in Applied Geophysics from Research Institute of Petroleum Exploration \& Development, Beijing, China. Recently he received a PhD degree in Geosciences from the University of Texas at Austin, Austin TX. He is specialized in seismic data analysis, machine learning and inverse problems.
\end{IEEEbiography}

\begin{IEEEbiography}
[{\includegraphics[width=1in,height=1.25in,clip,keepaspectratio]{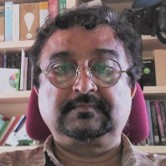}}]
{Mrinal K. Sen}
is a professor of Geophysics and holder of the Shell Companies Foundations Centennial Chair in Geophysics at the Department of Earth and Planetary Sciences, and the Institute for Geophysics at the University of Texas at Austin. He also serves as the director of the “Machine Learning and Data Analytics in Geosciences” graduate certificate program at the Jackson school. Mrinal received his B.SC-M.Sc degree from IIT(ISM) Dhanbad and PhD from the University of Hawaii at Manoa, USA. He has received many awards including the Honorary membership of the Society of Exploration Geophysicists “for extraordinary contributions as a geophysicist, educator and author”, the ‘Joseph C. Walter award for research excellence’, the ‘distinguished educator award’ at the University of Texas, and the distinguished alumnus award from IIT and the University of Hawaii at Manoa. He is the recipient of 2018 Virgil Kauffman gold medal of the Society of Exploration Geophysicists for making significant advancement in the sciences of exploration geophysics in the last five years. He has been chosen by the SEG to be the instructor of a one-day ‘Distinguished Instructor Short Course’ in 2024-25. 
\end{IEEEbiography}

\begin{IEEEbiography}
[{\includegraphics[width=1in,height=1.25in,clip,keepaspectratio]{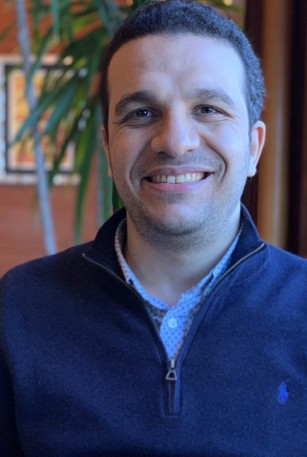}}]
{Abdelrahman Elmeliegy}
is a Computational Scientist at Idaho National Laboratory specializing in physics-based modeling, inverse problems, and AI/ML for nondestructive evaluation of nuclear materials. Previously, he was a Postdoctoral Fellow at UT Austin's Institute for Geophysics. He holds Ph.D. in Computational Inverse Problems and M.S. in Applied Mathematics both from NC State University in addition to M.S. in Structural Engineering, and B.S. in Civil Engineering both from Cairo University.
\end{IEEEbiography}

\begin{IEEEbiography}
[{\includegraphics[width=1in,height=1.25in,clip,keepaspectratio]{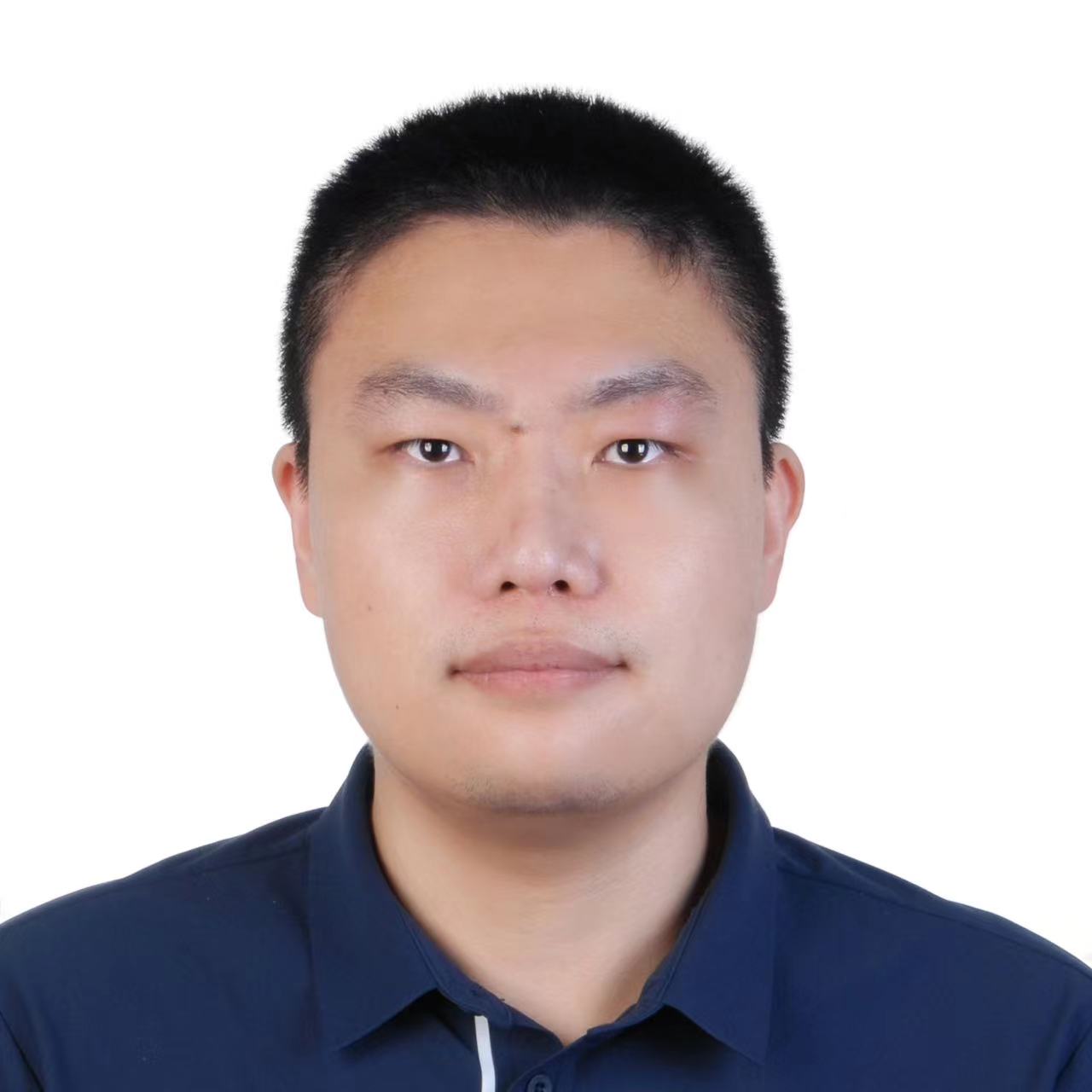}}]
{Shuo Zhang}
received the B.S. degree in Prospecting Technology and Engineering from China University of Petroleum (Beijing), Beijing, China in 2016, and the M.S. and Ph.D. degrees in Geosciences from The University of Texas at Dallas (UT Dallas), Richardson, USA, in 2018 and 2023, respectively.
Since 2023, he has been a Post-doctoral researcher in the Institute for Geophysics, The University of Texas at Austin, Austin, USA. His research interests include full waveform inversion, passive seismic imaging and monitoring.
\end{IEEEbiography}







\end{document}